\documentclass[final,3p,times,twocolumn,sort&compress]{elsarticle}
\usepackage[T1]{fontenc}
\usepackage[utf8]{inputenc}
\usepackage{amsmath}
\usepackage{amsfonts}
\usepackage{amssymb}
\usepackage{graphicx}
\usepackage{subfigure}
\usepackage{enumitem}
\usepackage{tabularx}
\usepackage{color}

\newcommand{\ie}{\textit{i.e.~}}
\newcommand{\eg}{\textit{e.g.~}}
\newcommand{\D}{\mathcal{D}}
\newcommand{\Na}{\mathcal{N}_a}
\newcommand{\Nc}{\mathcal{N}_c}
\newcommand{\Nt}{\mathcal{N}_t}

\newenvironment{definition}[1][Definition]{\begin{trivlist}
\item[\hskip \labelsep {\bfseries #1}]}{\end{trivlist}}

\newcolumntype{Y}{>{\centering\arraybackslash}X}
\newcommand{\revision}[1]{#1}

\begin{document}

\begin{frontmatter}
\title{An efficient Cellular Potts Model algorithm that forbids cell fragmentation}
\author{Marc Durand}
\ead{marc.durand@univ-paris-diderot.fr}
\author{Etienne Guesnet}

\address{Mati\`{e}re et Syst\`{e}mes Complexes (MSC), UMR 7057 CNRS \& Universit\'{e} Paris Diderot, 10 rue Alice Domon et L\'{e}onie Duquet, 75205 Paris Cedex 13, France.}
\date{\today}
\begin{abstract}
The Cellular Potts Model (CPM) is a lattice based modeling technique which is widely used for simulating cellular patterns such as foams or biological tissues. Despite its realism and generality, the
standard Monte Carlo algorithm used in the scientific literature to evolve this model \revision{preserves connectivity of cells on a limited range of simulation temperature only.} We present a new algorithm in which cell fragmentation is forbidden for all simulation temperatures. This allows to significantly enhance realism of the simulated patterns. It also increases the computational efficiency compared with the standard CPM algorithm even at same simulation temperature, thanks to the time spared in not doing unrealistic moves. Moreover, our algorithm restores the detailed balance equation, ensuring that the long-term stage is independent of the chosen acceptance rate and chosen path in the temperature space.
\end{abstract}


\end{frontmatter}

\section{Introduction}
The Cellular Potts Model (CPM) -- or Glazier-Graner-Hogeweg model -- developed by Glazier and Graner \cite{graner_simulation_1992,glazier_simulation_1993,glazier_magnetization_2007} has become a common technique to simulate cellular patterns either in physics or biology.
%
Originally, the CPM was proposed to test the \textit{differential adhesion hypothesis} suggested by Steinberg \cite{Steinberg_1963} to explain the spontaneous segregation of cells of different type, a phenomenon known as \textit{cell sorting}.
The CPM approach describes cellular systems based on the following assumptions: \textit{i)} cells are spatially extended but internally structureless objects; \textit{ii)} cells and associated fields are discretized onto a lattice; \textit{iii)} it describes most cell behaviors and interactions in terms of an effective energy or Hamiltonian $\mathcal{H}$;
\textit{iv)} the classic implementation of the CPM employs a modified Metropolis Monte-Carlo algorithm which chooses update sites randomly and accepts them with a Metropolis probability. A \textit{simulation temperature} $T$ then determines the probability of a configuration. For thermal physical systems, $T$ is the actual temperature (up to Boltzmann constant $k_B$). For biological cells, actual temperature is too low to induce significant fluctuations, and $T$ simulates the membrane fluctuations due to cell activity \cite{mombach_quantitative_1995}.  

In its original form \cite{graner_simulation_1992, glazier_simulation_1993}, the CPM Hamiltonian $\mathcal{H}$ has only two contributions: a boundary term and a compressive term, which write, for a two-dimensional pattern:
\begin{equation}
\mathcal{H}=\sum_{\substack{neighboring \\ sites  \langle k,l \rangle}}J_{\tau,\tau^\prime}\left(1-\delta_{\sigma_k,\sigma_l}\right)+\frac{B}{2A_0}\sum_{\substack{cells \\ i}}\left(A_i-A_0\right)^2.
\label{Hamiltonian}
\end{equation}
Here $\sigma_k$ and $\sigma_l$ are the site values of site $k$ and $l$, respectively. \revision{A cell $i$ consists of all sites in the lattice with site value $i$.} $\delta$ is the Kronecker delta symbol: $\delta_{m,n}=1$ if $m=n$, and $0$ otherwise. $\tau$ and $\tau^\prime$ are abbreviations for $\tau(\sigma_k)$ and $\tau(\sigma_l)$, the cell types which can be attributed to cells with respective value $\sigma_k$ and $\sigma_l$. $J_{\tau,\tau^\prime}\left(=J_{\tau^\prime,\tau}\right)$ is the energy per unit contact length between cell types $\tau$ and $\tau^\prime$. For a foam, there is a unique cell type and hence $J$ reduces to a constant proportional to the surface tension. $B$ is the bulk modulus of the internal fluids, \revision{$A_i$ is the area of cell $i$,} and $A_0$ the \textit{target area}, \textit{i.e.} the area that the enclosed fluid would occupy for its pressure to be equal to the surrounding pressure. 
The first sum in Eq. (\ref{Hamiltonian}) is carried over neighbouring sites $\langle k,l \rangle$ and represents the boundary energy: each pair of neighbours having unmatching indices determines a boundary and contributes to the boundary energy.
The second sum in Eq. (\ref{Hamiltonian}), carried over all the cells that constitute the pattern, is the compressive energy of the cells. 
Since then, additional terms have been added to the Hamiltonian to account for real cell behaviors, or application of external fields \cite{ouchi_improving_2003,glazier_magnetization_2007,magno_biophysical_2015}.

Because of its flexibility, extensibility and ease of use, the CPM has been widely and successfully used in different domains of physics, biology or \revision{medicine} \cite{graner_simulation_1992,glazier_simulation_1993,mombach_quantitative_1995,
ouchi_improving_2003,mombach_rounding_2005,kafer_cell_2007,oates_quantitative_2009,ariotti_tissue-resident_2012,fortuna_growth_2012,Albert2014,magno_biophysical_2015,Kabla3268,Hallou2016}.
In order to prevent the apparition of heterogeneous sites (\textit{spontaneous nucleation}), and hence preserve the connectivity of the cells, the CPM uses a modified Metropolis algorithm (MMA), in which a lattice site is authorized to be changed in one of its neighboring values only.
Although this algorithm substantially improves the realism of the simulated patterns, it has yet some flaws.
First, it works on a limited range of temperature: if the temperature is too low, the system is trapped in one of its many energetic valleys.
If the temperature is too high, connectivity of cells is no more guaranteed: boundaries can become highly contorted, provoking the detachment of small (usually single lattice site) fragments. 
This upper limit on the usable range of temperature inevitably makes simulations very slow if spatial resolution is not chosen with care \cite{magno_biophysical_2015}. 
To overcome this limitation, sophisticated parallelization techniques have been developed, with varying degrees of success \cite{gusatto_efficient_2005,chen_parallel_2007,tapia_parallelizing_2011}. 

Second, the MMA --- unlike the classic Metropolis algorithm --- does not satisfy the detailed balance condition \cite{glazier_magnetization_2007,Zajac_2002,voss-bohme_multi-scale_2012}. Detailed balance ensures that the long term distribution of configurations obeys Boltzmann statistics.
This is critical when intending to study the equilibrium configuration of thermal systems (\textit{e.g.} thermally shuffled foam). 
When CPM is used as a purely kinetic model
for out-of-equilibrium systems (\textit{e.g.} coarsening foams, morphogenesis, cell sorting), detailed balance condition is irrelevant.
However, using algorithms that satisfy detailed balance can still be of interest for such systems, as it guarantees that the long-term stage eventually reached by the simulation depends only on the final simulation temperature, and not on the chosen path in the temperature space or the chosen acceptance rate (\textit{e.g.} Metropolis, heat-bath, Glauber \cite{newman_monte_1999}).

In this paper, we present a new algorithm in which both fragmentation and spontaneous nucleation are forbidden.  Normally, the inspection of cell connectivity is computationally very expensive: at every modification of a site value, one must check that there is always a path connecting any pair of lattice sites that belong to the same cell.
However, we show that a test of  \textit{local connectivity} in the neighborhood of the modified site is necessary and sufficient for ensuring that cells remain \textit{simply} connected (\ie homeomorphic to a disk): fragmentation or handle formation are then prohibited. 
Hence, this new algorithm enhances realism for most cellular systems, at least in non-pathological situations.
At given simulation temperature, these modifications increase the computational efficiency, as time used for local connectivity inspection is largely compensated by the time not wasted in unrealistic cell fragmentation moves. Moreover, this new algorithm allows to increase the simulation temperature, thereby speeding up simulation time, while preserving connectivity of the cells. Finally, our proposed algorithm restores the condition of detailed balance at all temperatures. 
Our algorithm is much simpler to implement than parallel algorithms. Nevertheless, both techniques can be combined together to simulate larger systems.

The structure of the paper is as follows: in Section \ref{Modified_algo}, we detail the standard algorithm used in most CPM simulations, and emphasize that it does not prevent from cell fragmentation or handle formation, and violates detailed balance condition. We also point out the three distinct notions of neighborhood that are used in CPM. In Section \ref{section_NMA}, we detail a new algorithm that resolves these issues. The notion of local connectivity, on which this algorithm relies, is properly defined.
In Section \ref{section_optimization}, we show that a careful choice for the different neighborhoods makes the test of local connectivity very fast. In Section \ref{section_benchmark}, we \revision{present} cell sorting simulations on a two-dimensional square lattice and compare the efficiency of the two algorithms. The new algorithm performs better for all tested temperatures: the convergence to the long-term, steady stage is computationally more efficient for a same simulation temperature, and the long-term stage \revision{obtained} after smoothing (``annealing'') procedure is independent of the temperature value.
Finally, in Section \ref{Outlook}, we discuss extension of the algorithm to other 2D or 3D lattices.

\section{Classical CPM algorithm: Modified Metropolis Algorithm (MMA)}  \label{Modified_algo}
\subsection{Motivations}
Algorithm used in CPM simulations is adapted from Metropolis algorithm, which is one of the simplest and most popular algorithm used in Monte Carlo simulations. Applied to the Potts model \cite{newman_monte_1999, glazier_magnetization_2007}, in which each site, or ``spin'', can have $Q$ different values, the  Metropolis algorithm consists of the following steps:
\begin{enumerate}
\item Randomly select a lattice site $i$. Call this site the \textit{candidate site}. Let $\sigma$ be its value.
\item Randomly select a value $\sigma^\prime$ among the $Q$ possible site values. Call this the \textit{target value}.
\item Calculate the change in energy $\Delta E$ resulting in changing the site value from $\sigma$ to $\sigma^\prime$.
\item Accept the site value modification with probability $\mathcal{A}(\sigma \rightarrow \sigma^\prime)=\min(1,e^{- \Delta E/T})$.
\item Increment the number of copy attempts and go back to step 1.
\end{enumerate}

This algorithm allows \textit{spontaneous nucleation}, \textit{i.e} the appearance of site value different from its neighborhood. This of course does not realistically simulate cellular systems, in which every cell must be (simply) connected. Metropolis algorithm must be adapted in order to eliminate nucleation. The algorithm that has been used by Graner \& Glazier \cite{graner_simulation_1992,glazier_simulation_1993} (and since then in most of the scientific literature on CPM) replaces step 2. with:
\begin{enumerate}[label=\arabic*b.]
\setcounter{enumi}{1}
\item Randomly select a site from the candidate site's neighbor list. Call this the \textit{target site}, and let $\sigma^\prime$ be its value.
\end{enumerate}
Note that here a site, and not a site value, is randomly selected.
This algorithm is commonly named \textit{Modified Metropolis Algorithm} (MMA). The primary reason to use the Metropolis acceptance ratio $\mathcal{A}(\sigma \rightarrow \sigma^\prime)$ in step (4.) is that the average time evolution of the configuration then obeys the \textit{Aristotelian} or \textit{overdamped} force-velocity relation \cite{glazier_magnetization_2007}: $\mathbf{v} \propto \boldsymbol{\nabla} \mathcal{H}$. Nevertheless, the acceptance ratio of any other single-spin-flip algorithm (\eg Glauber, Heat Bath \cite{newman_monte_1999}) could be used; in any case, the modification from step (2.) to step (2b.) is required to suppress spontaneous nucleations.

\subsection{Neighborhoods} \label{section_neighborhoods}
It can be noticed that the algorithm introduces two distinct notions of neighborhood so far: the first one, which we shall call \textit{coupling neighborhood}, is the one which is used in the first sum of the Hamiltonian (\ref{Hamiltonian}) to calculate the interface energy. We note it $\Nc$.
The second one, that we shall call \textit{target neighborhood}, is the one used in step (2b), and represents the set $\Nt$ of lattice sites \revision{from which the target value is chosen.} 
\revision{Actually, a} third neighborhood is used in CPM simulations, when calculating the number of sides (or more exactly, of neighboring cells) of every cell. We shall call it \textit{adjacency neighborhood} $\Na$: for every site of a cell, we list the site values that are in this neighborhood. The total list of different values provides the list of neighboring cells.

Distinction between these three neighborhoods is often disregarded in literature. Yet, they can be chosen independently; a careful choice can tremendously increase the efficiency of the algorithm, as will be discussed in Sect. \ref{section_optimization}. 

\subsection{Limitations of the MMA}
\begin{figure}[htb]
\begin{center}
\includegraphics[width=0.9\columnwidth]{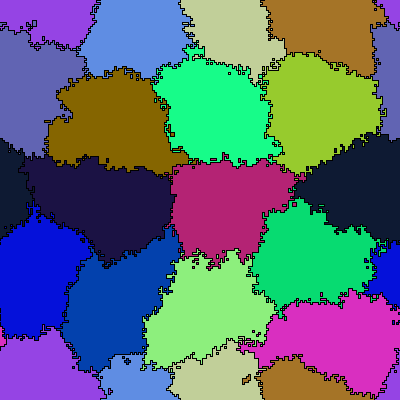}
\caption{Cell fragmentation observed with the MMA at high temperature values: fragments of one or more lattice sites are surrounded by mismatched site values. Here each cell has a different color to highlight the presence of fragments.}
\label{fig:fragments}
\end{center}
\end{figure}

\begin{figure}[htb]
\centering
\subfigure[]{
\includegraphics[width=0.2\columnwidth]{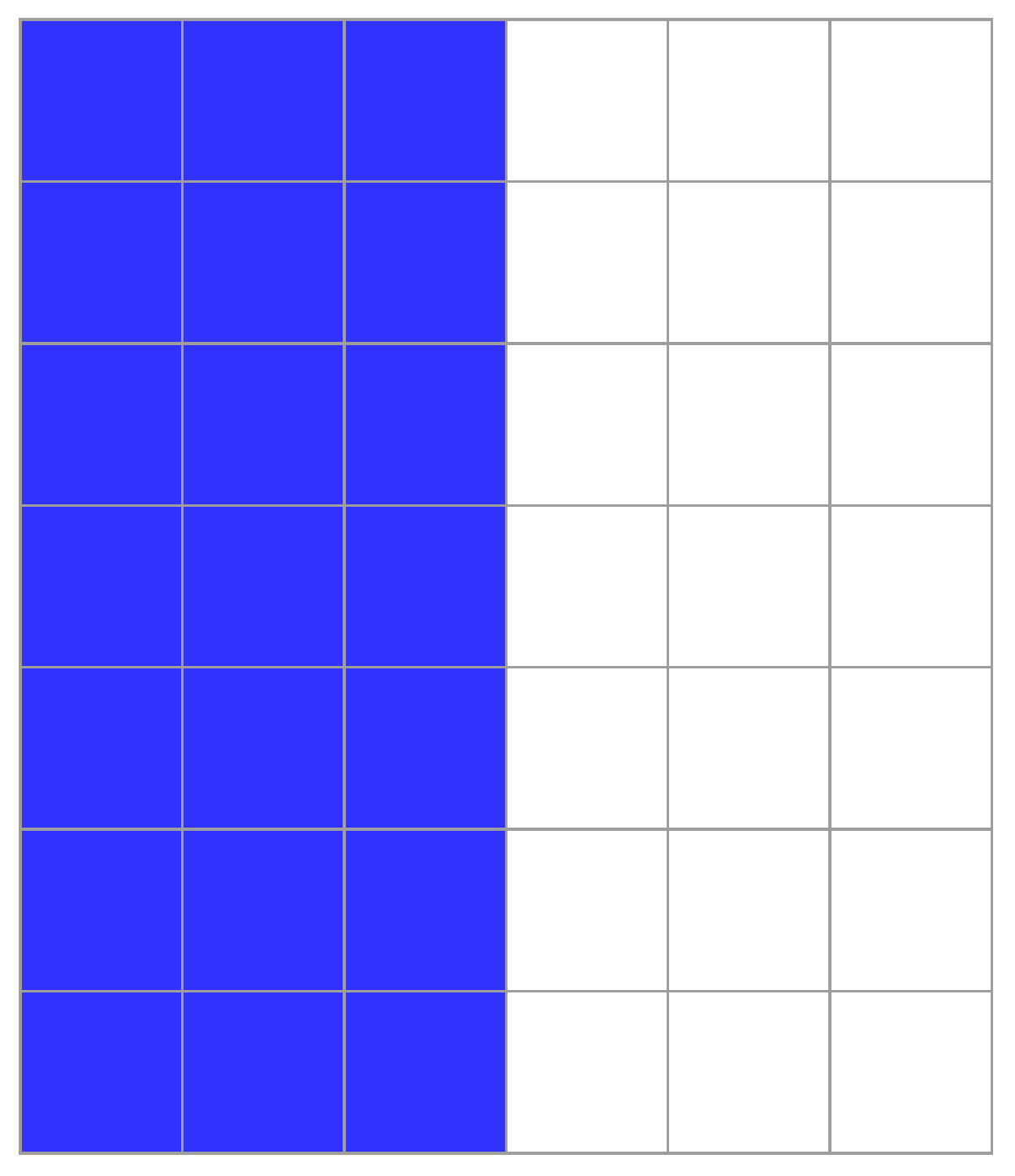}}
\hspace{0.03\columnwidth}
\subfigure[]{
\includegraphics[width=0.2\columnwidth]{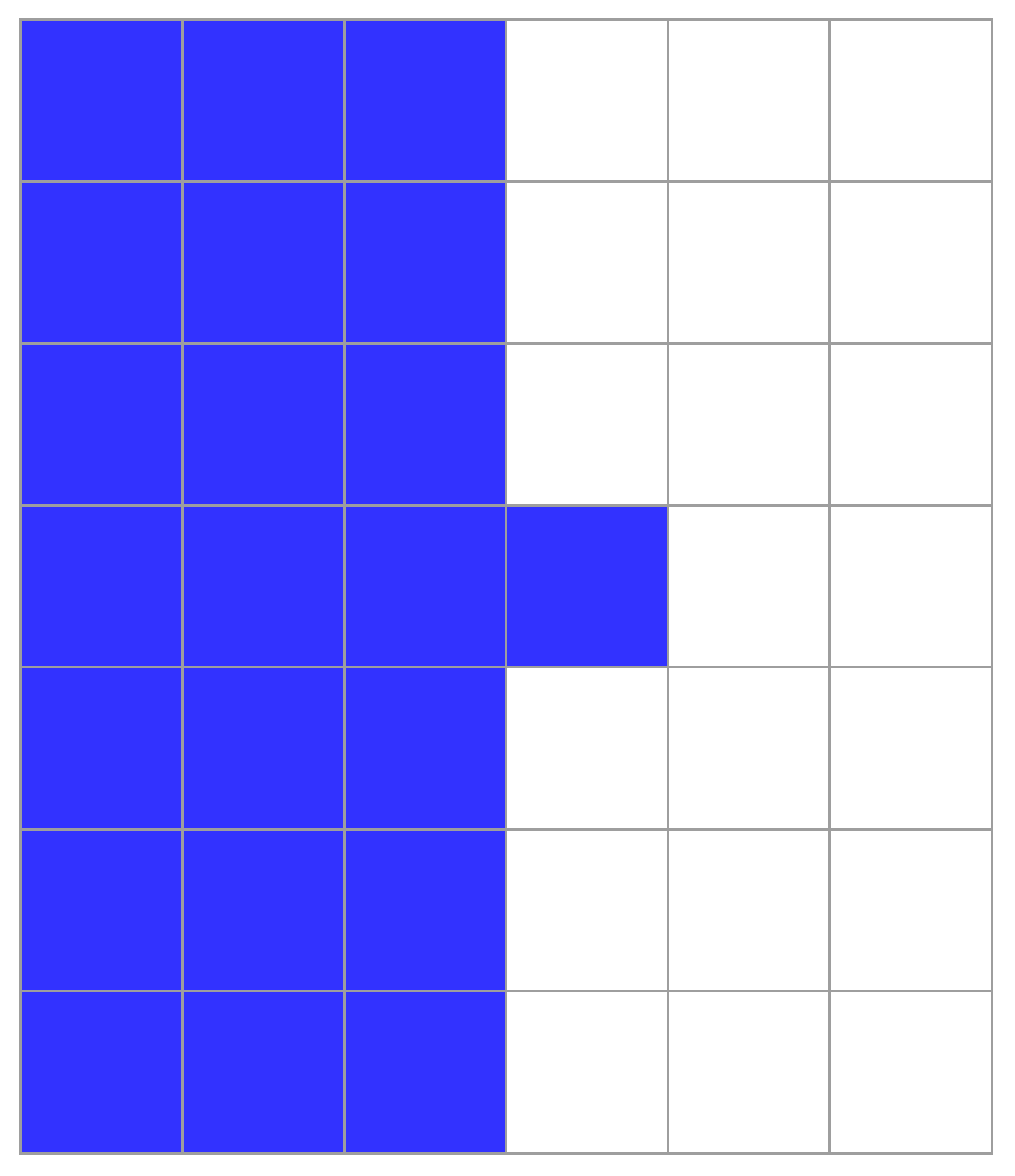}}
\hspace{0.03\columnwidth}
\subfigure[]{
\includegraphics[width=0.2\columnwidth]{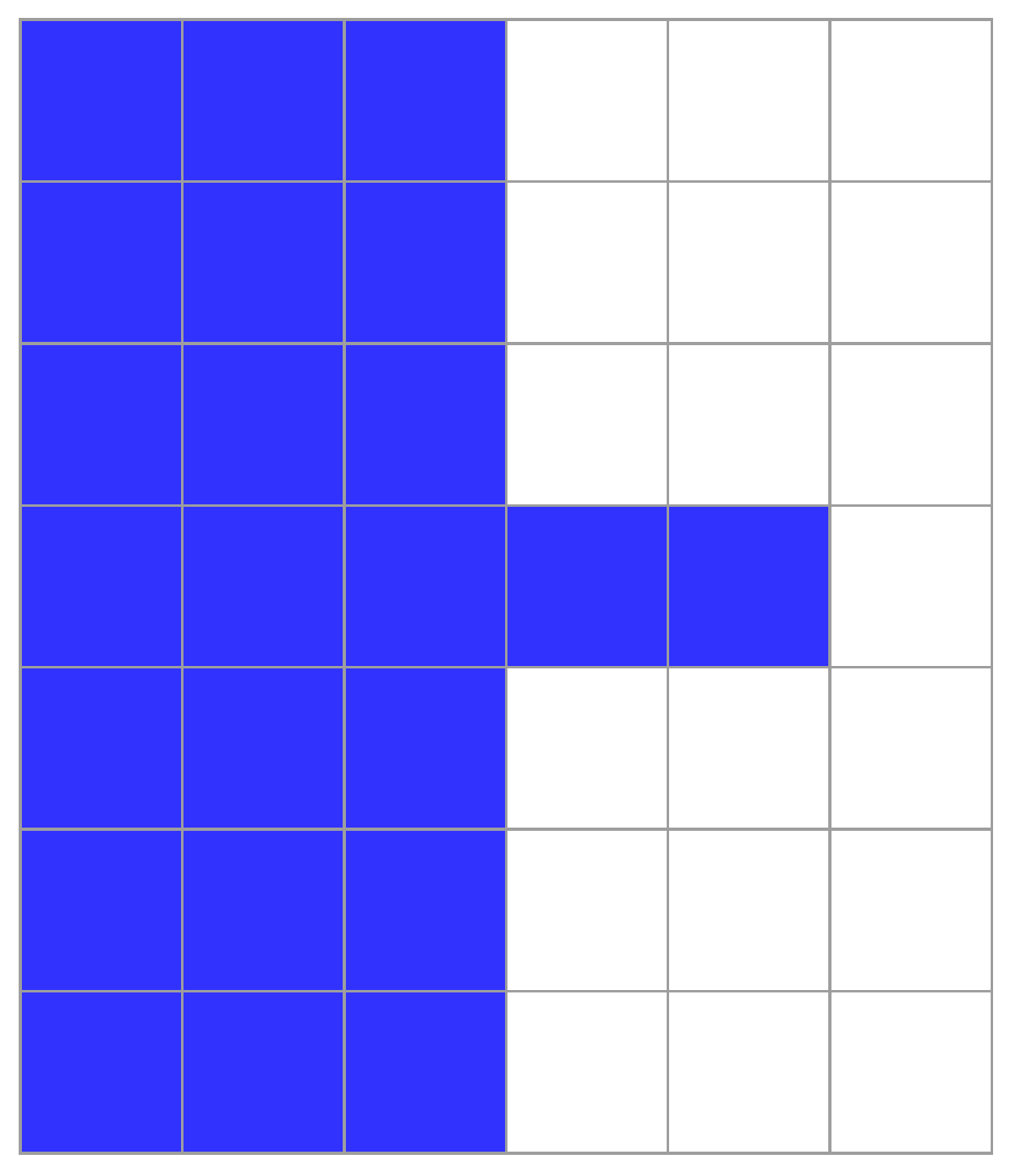}}
\hspace{0.03\columnwidth}
\subfigure[]{
\includegraphics[width=0.2\columnwidth]{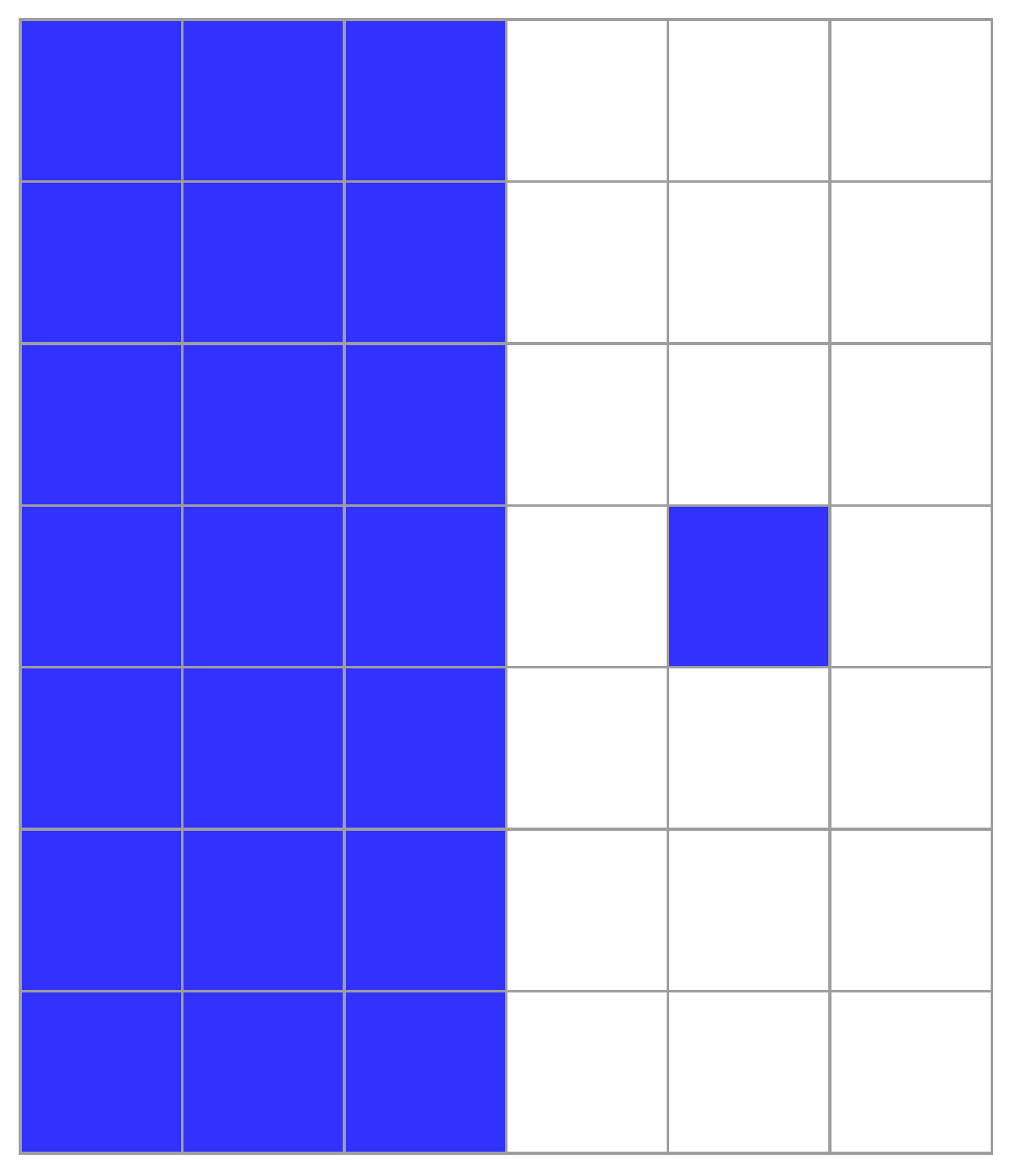}}
\caption{Simplest scenario of one-pixel-size fragmentation resulting in an isolated pixel.
}
\label{fragmentation}
\end{figure}

In addition to improving realism, the modification of step (2.) to step (2b.) speeds up the computing time by reducing the number of possible target values. But it has still some limitations. First, although spontaneous nucleation is forbidden, heterogeneous sites (\ie sites with mismatched neighboring sites) still appear, as illustrated in Fig. \ref{fig:fragments}: they result from the detachment from the interface of fragments whose size is eventually reduced to a unique lattice site. The number and size of fragments increase with simulation temperature $T$, as interfaces get more and more crumpled. 
Simplest scenario of fragmentation is shown in Fig. \ref{fragmentation}: a fragment of one lattice site size detaches from the interface in a three steps process. \revision{To create such a fragment, the system must cross the energy barrier $\simeq z_c J$ -- corresponding to the transition from configuration \ref{fragmentation}(a) to configuration \ref{fragmentation}(c) -- where $z_c$ is the number of sites in the coupling neighborhood $\Nc$ and $J$ the energy per unit contact length. It is therefore less than $z_c J \ell$, the typical energy barrier which must be crossed to trigger a neighbor switching between 4 cells with characteristic edge length $\ell$ \cite{Durand_PRL}. Hence, fragmentation appears at temperatures below the temperature required to initiate cell diffusion.}

Like spontaneous nucleation, fragmentation is undesirable because it does not mimic realistically cellular systems, at least in normal (non pathological) situations. Moreover, it causes an overestimation of the number of sides of the cells and complicates the calculation of interfacial energy. Fragmentation is also undesirable in measuring cell mean square displacement because it may cause the center of mass of a domain to jump a large distance after a single spin flip.
Therefore, CPM simulations must be limited to temperature values low enough for the fragments to disappear with a rate at least equal to the rate at which they appear. This low temperature range is quite restrictive, as the systems which are simulated with CPM usually have a rough energy landscape: they are trapped in successive local minima and thus evolve slowly.
To reduce the visual presence of such fragments, ``annealing'' steps at vanishing temperature are often performed before measurements \cite{graner_simulation_1992, glazier_simulation_1993, glazier_magnetization_2007}. However, fragmentation still occurs between annealing steps, and has a strong effect on the kinetics of the system. In fact, even the long term stage of the simulation can be affected by the fragmentation, despite the annealing procedure, as will be shown in Sect. \ref{section_benchmark}.



Another peculiarity of the MMA is that by changing step (2.) to step (2b.), detailed balance condition is not satisfied anymore.
In its canonical formulation, the detailed balance equation reads:
\begin{equation}
%
\dfrac{g(x_i:\sigma\rightarrow\sigma^\prime) \mathcal{A}(x_i:\sigma\rightarrow\sigma^\prime)}{g(x_i:\sigma^\prime\rightarrow\sigma) \mathcal{A}(x_i:\sigma^\prime\rightarrow\sigma)}=e^{- \Delta E/T},
\label{detailed_balance}
\end{equation}
where $g(x_i:\sigma\rightarrow\sigma^\prime)$ is the \textit{selection probability}, which is the probability, given an initial state of the system, that our algorithm will generate a new target state that differs from the initial state by changing the value of one single site $i$ from $\sigma$ to $\sigma^\prime$. Detailed balance criterion, or microreversibility, guarantees that the long term distribution of configurations follows the Boltzmann statistics \cite{newman_monte_1999}. Hence, detailed balance condition must be respected to adequately simulate systems at thermodynamic equilibrium. 
\revision{Detailed balance condition is irrelevant when simulating out-of-thermodynamic-equilibrium systems such as biological systems or coarsening foams;} the primary requested feature of the algorithm is that it mimics realistically the time evolution of the system \cite{glazier_magnetization_2007}. However, various different algorithms usually do satisfy this request. Since the CPM approach follows classic Monte Carlo schemes, the use of an algorithm that also satisfies detailed balance criterion ensures that a steady state exists (\eg long term stage of cell sorting), and is the same for any specific acceptance rate that satisfies detailed balance condition. 
Moreover, microreversibility solves the degenerate long-term behavior associated with the MMA \cite{voss-bohme_multi-scale_2012}.

Violation of detailed balance is noticeable in particular in the case of spontaneous nucleation: although spontaneous nucleation is forbidden (thanks to the modification of step 2 to step 2b), its reverse process -- that is, the disappearance of a heterogeneous site -- is allowed, and even wanted to avoid their proliferation, resulting from cell fragmentation. We could (naively) think of restoring detailed balance by forbidding the disappearance of heterogeneous sites. However, this approach generates a proliferation of nucleated sites as temperature increases, the only way for an isolated site to disappear is by reaching the cell it belongs to. This would certainly not improve the realism of the simulations. Moreover, that would not restore microreversibility in all situations: detailed balance condition also requires a modification of step (2b.), as this will appear below \revision{(see also the Appendix)}.


\section{Connectivity Algorithm (CA)} \label{section_NMA}





It is clear from the discussion above that an algorithm that would forbid cell fragmentation, in addition to spontaneous nucleation, would preserve the connectivity of the cells, and then would solve the limitations inherent to the MMA.
Two cells are possibly affected by the modification of a site value: the \emph{candidate cell} and the \emph{target cell}, defined as the cells whose the candidate and the target lattice site belong to, respectively. Prior to the modification of the candidate site value, one must check that these two cells will both remain \textit{connected}. In fact, by choosing $\Nt \subseteq \Na$, we are ensured that the target cell stays connected, by construction. Still, the inspection of the connectivity of the candidate cell is very costly in computing time: for \textit{every} couple of lattice sites that belong to the cell, one must inspect that there is a \textit{path} that links them, where a path is defined as a list of lattice sites with same value, and each of which belongs to the adjacency neighborhood of the preceding one\footnote{For convenience, a path -- and hence the connectivity property -- is defined based on the adjacency neighborhood $\Na$, but this notion actually introduces a fourth notion of neighborhood that could have been chosen independently of the three neighborhoods already defined in Sect. \ref{section_neighborhoods}.}.

Instead, we propose to test the \textit{local connectivity} of the cell. To define this property, we need to introduce first the \textit{local connectivity domain} $\mathcal{D}_c(i)$, as being any set of lattice sites that contains the adjacency neighborhood of site $i$, but not site $i$ itself.
%
The local connectivity property is then defined as follows: 
\begin{definition}
a cell $\mathcal{C}$ is \textit{locally connected} at site $i$, within the local connectivity domain $\mathcal{D}_c(i)$, if and only if the sites of $\mathcal{C}$ that are in the adjacency neighborhood of $i$ are connected through paths that are included in $\mathcal{D}_c(i)$.
\end{definition}
Hence, if $\mathcal{C}$ is locally connected at position $i$, within the local connectivity domain $\mathcal{D}_c(i)$, we are ensured that there are paths connecting the sites of $\mathcal{C} \cap \Nt$ and which do \textit{not} contain site $i$. 

\begin{figure}[htb]
\centering
\includegraphics[width = 0.45\columnwidth]{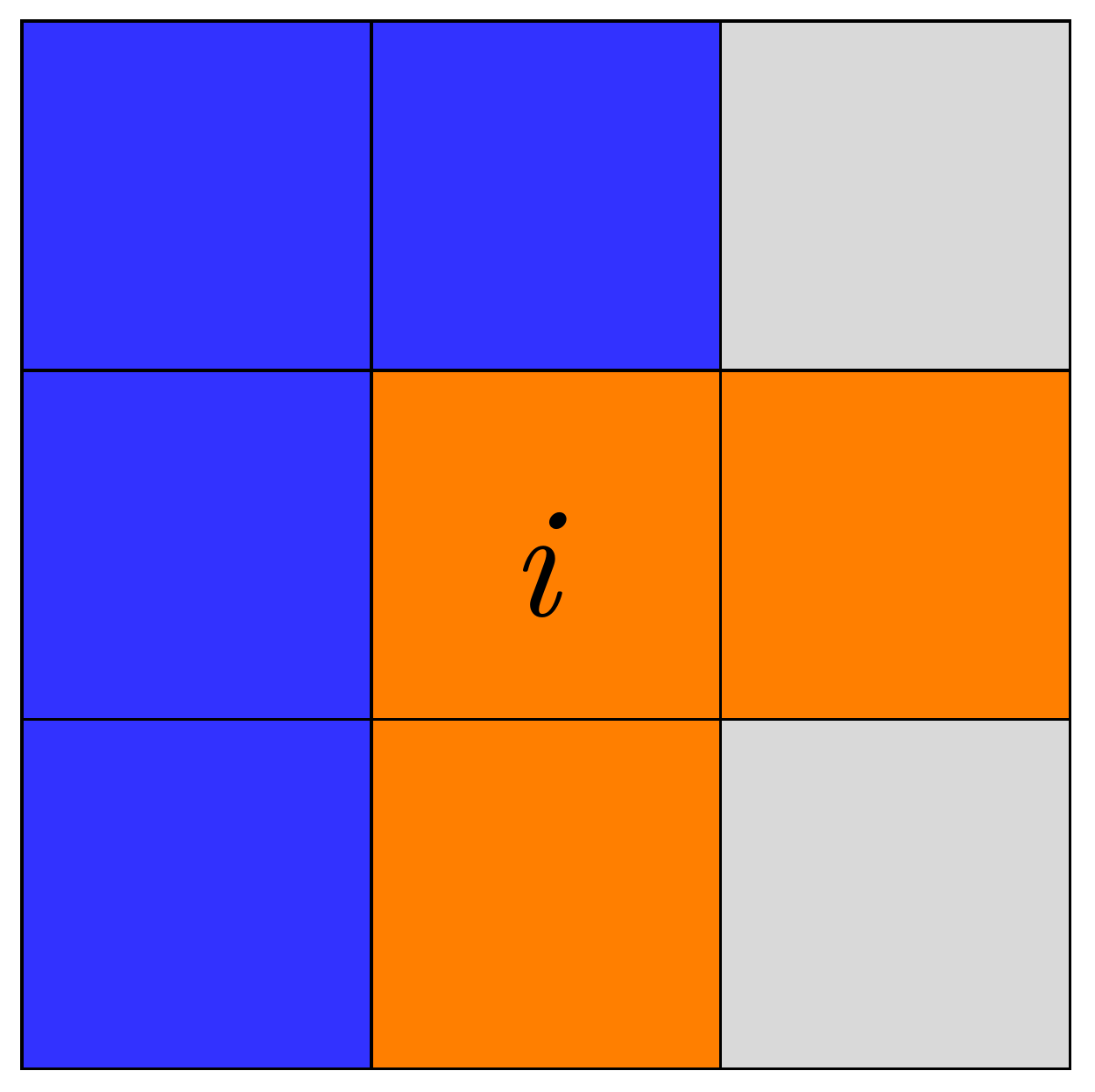}
\caption{Illustration of the notion of \textit{local connectivity}: suppose $\mathcal{D}_c(i)$ is composed of the eight sites that surround site $i$ and $\Na$ is composed of the four side-adjacent sites. Then, the blue cell is locally connected at site $i$, while the orange and grey cells are not.}
\label{local-connectivity-domain}
\end{figure}

Let us illustrate this notion of local connectivity with the example given in Fig. \ref{local-connectivity-domain}: let $\mathcal{D}_c(i)$ be composed of the eight sites that surround site $i$ (Moore neighborhood), and the connectivity neighborhood $\Nt$ be composed of the four side-adjacent sites (Von Neumann neighborhood). Then, the blue cell is locally connected at site $i$, while the orange and grey cells are not. This example emphasizes that a cell $\mathcal{C}$ can be locally connected at site $i$ even if $i$ is not in $\mathcal{C}$, \revision{and not locally connected at site $i$ even if $i$ is in $\mathcal{C}$.}  

The proposed test then consists, in a first step, to check the local connectivity of the candidate cell at the randomly selected site $i$ (candidate site), before eventually accepting the modification of its value.
This simplification substitutes a global test with a local one, resulting in a huge saving of computing time. However, is this local test a necessary and sufficient condition to guarantee the non-fragmentation of the cells ?

\begin{figure}[htb]
\centering
\subfigure[]{
	\label{one-pixel-thick}
    \includegraphics[width = 0.48\columnwidth]{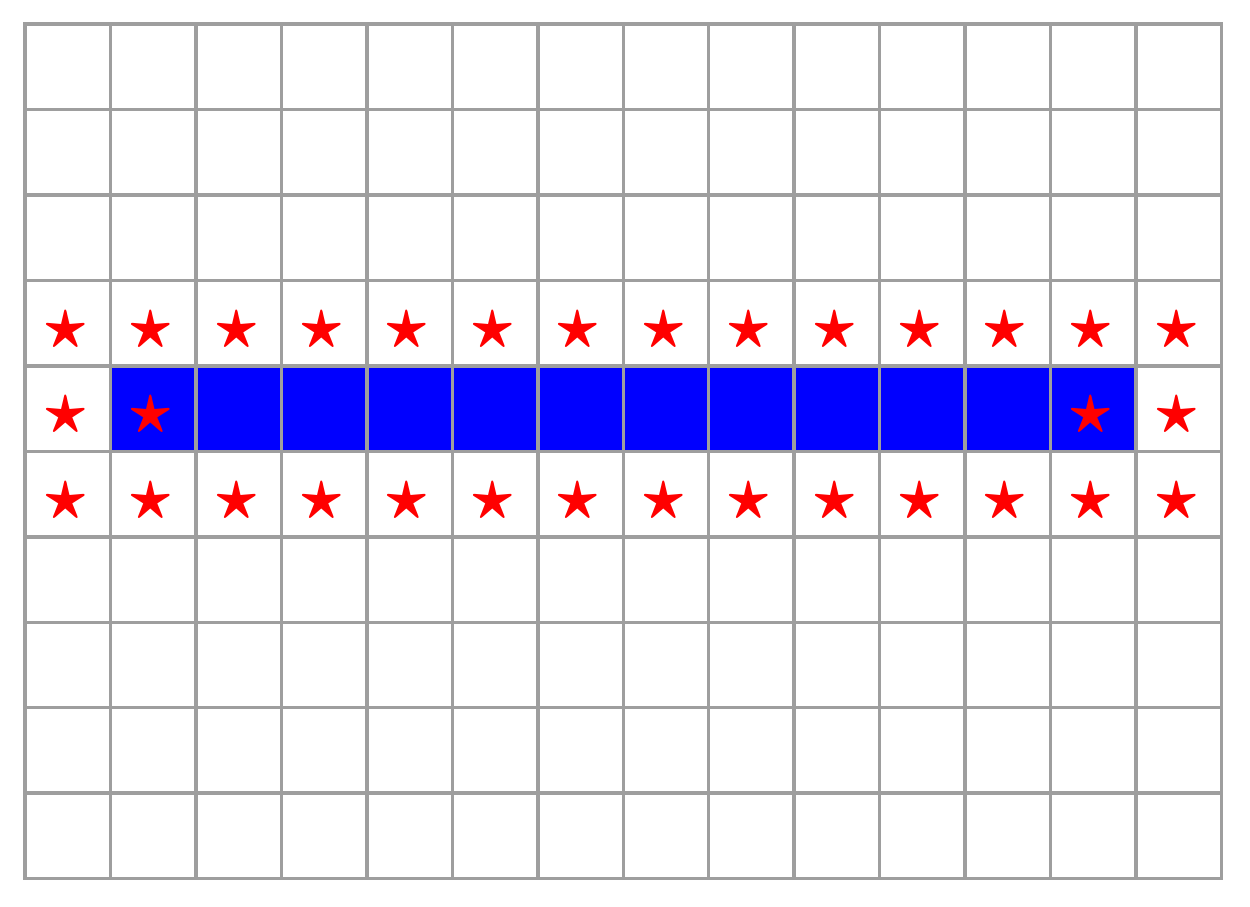}}
\subfigure[]{
	\label{non-connected-cell}
    \includegraphics[width = 0.48\columnwidth]{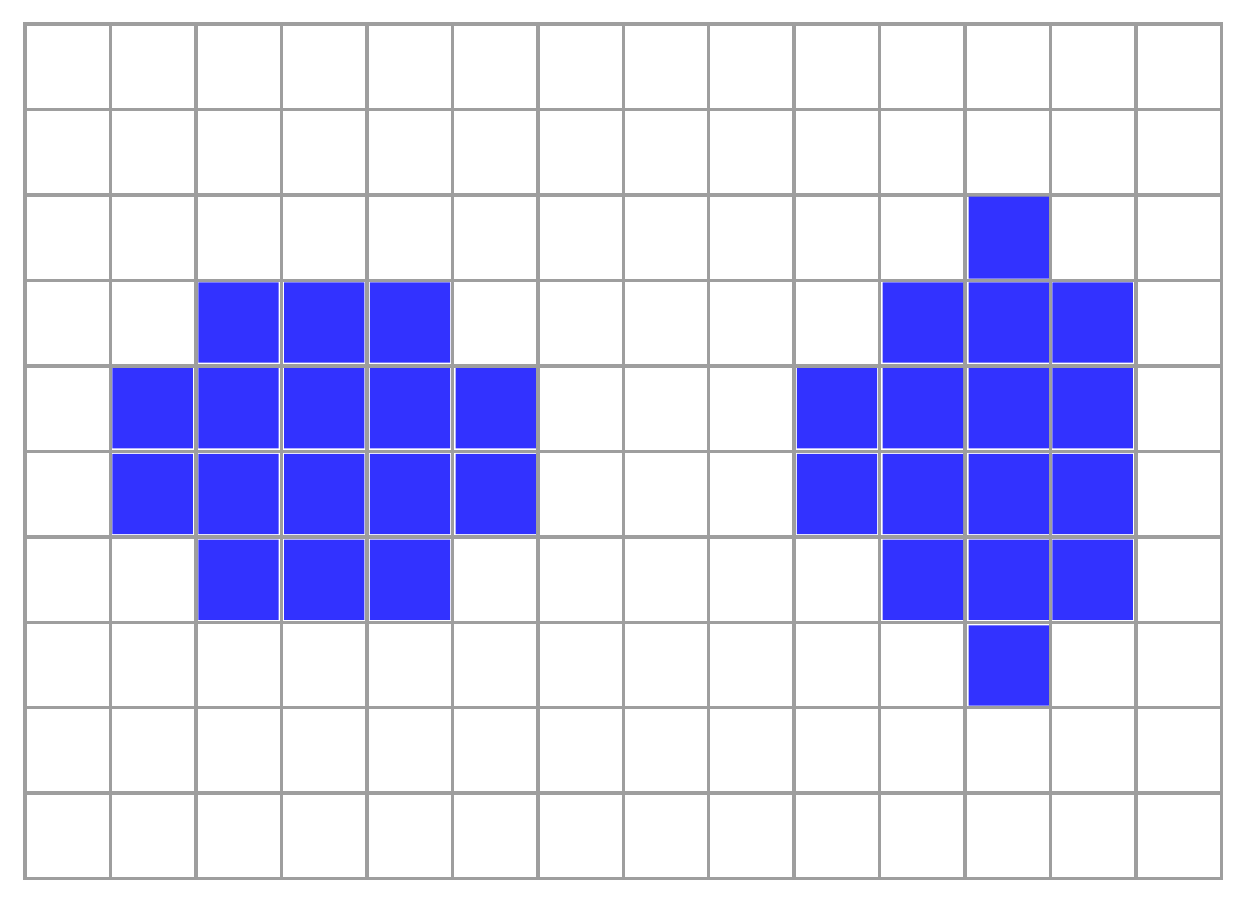}}
\caption{Illustration of the non-equivalence between connectivity and local connectivity: (a) a flat cell which is one-site thick is connected, yet it is not locally connected everywhere. Red stars indicate the lattice sites inside and outside the cell over which the cell is locally connected, with same $\D_c$ and $\Na$ as in Fig. \ref{local-connectivity-domain}. (b) a cell composed of two disjoint fragments is locally connected, yet not connected.}
\end{figure}

A connected cell is not necessarily locally connected everywhere: for instance, a flat cell with 1 lattice site thickness is not locally connected, excepted at its ends (see Fig. \ref{one-pixel-thick}). \revision{This is also the case of the orange cell in Fig. \ref{local-connectivity-domain}.} Conversely, a cell that is locally connected on everyone of its sites is not necessarily connected, as this is illustrated in Fig. \ref{non-connected-cell}, where a cell is divided into two fragments which are both locally connected.
Nevertheless, if we ensure that, at the initial time (cellular pattern formation), cells are connected, then the local connectivity test will prevent cell fragmentation at any subsequent time. Local connectivity test at site $i$ is then a sufficient condition for the candidate cell to remain connected\footnote{A common way to generate the cellular pattern is from the growing of cell ``seeds'' in a medium, which is treated as a special cell without area constraint. One then must be careful to momentarily allow the fragmentation of the medium, until cells touch each others, so that the removing of medium between the cells is achieved in a small amount of time.}. 

\begin{figure}[htb]
\centering
\subfigure[]{
    \includegraphics[width = 0.48\columnwidth]{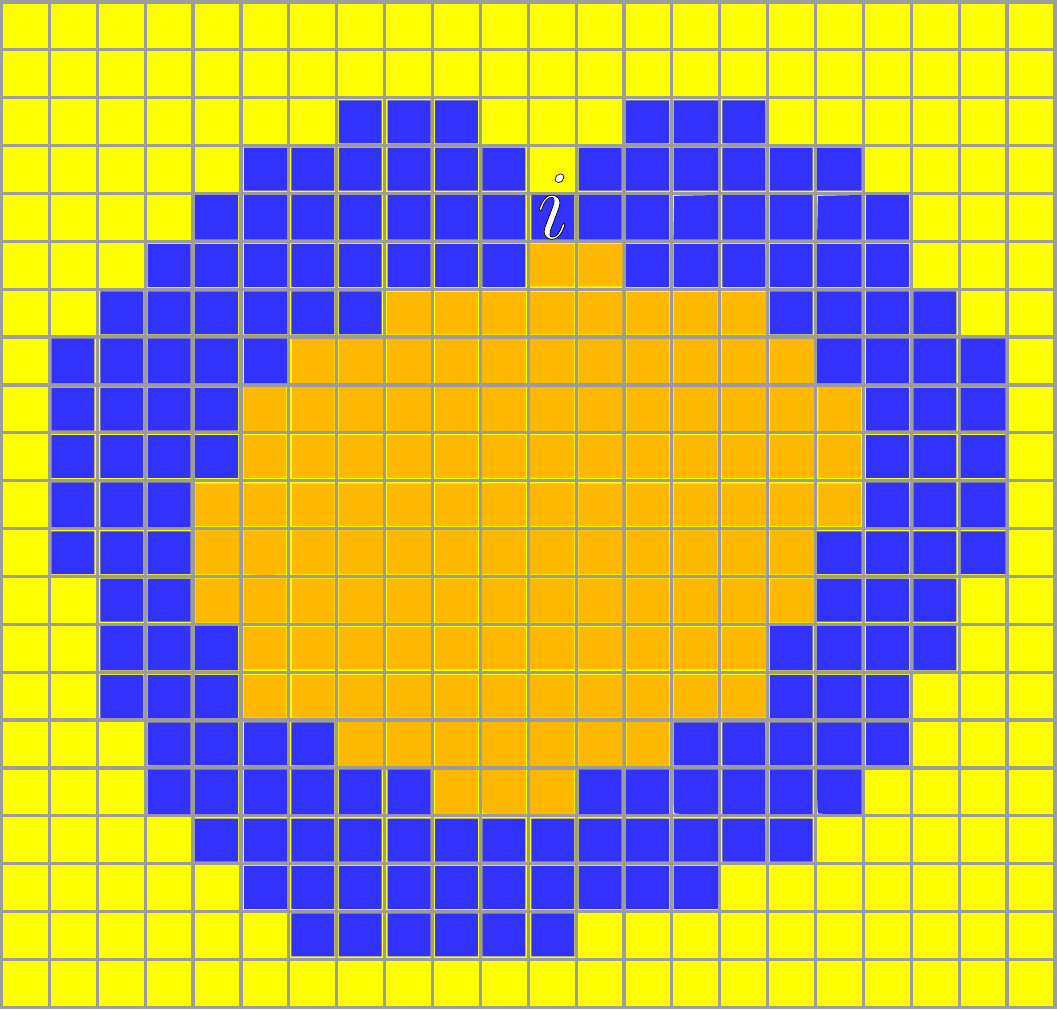}}
\subfigure[]{
    \includegraphics[width = 0.48\columnwidth]{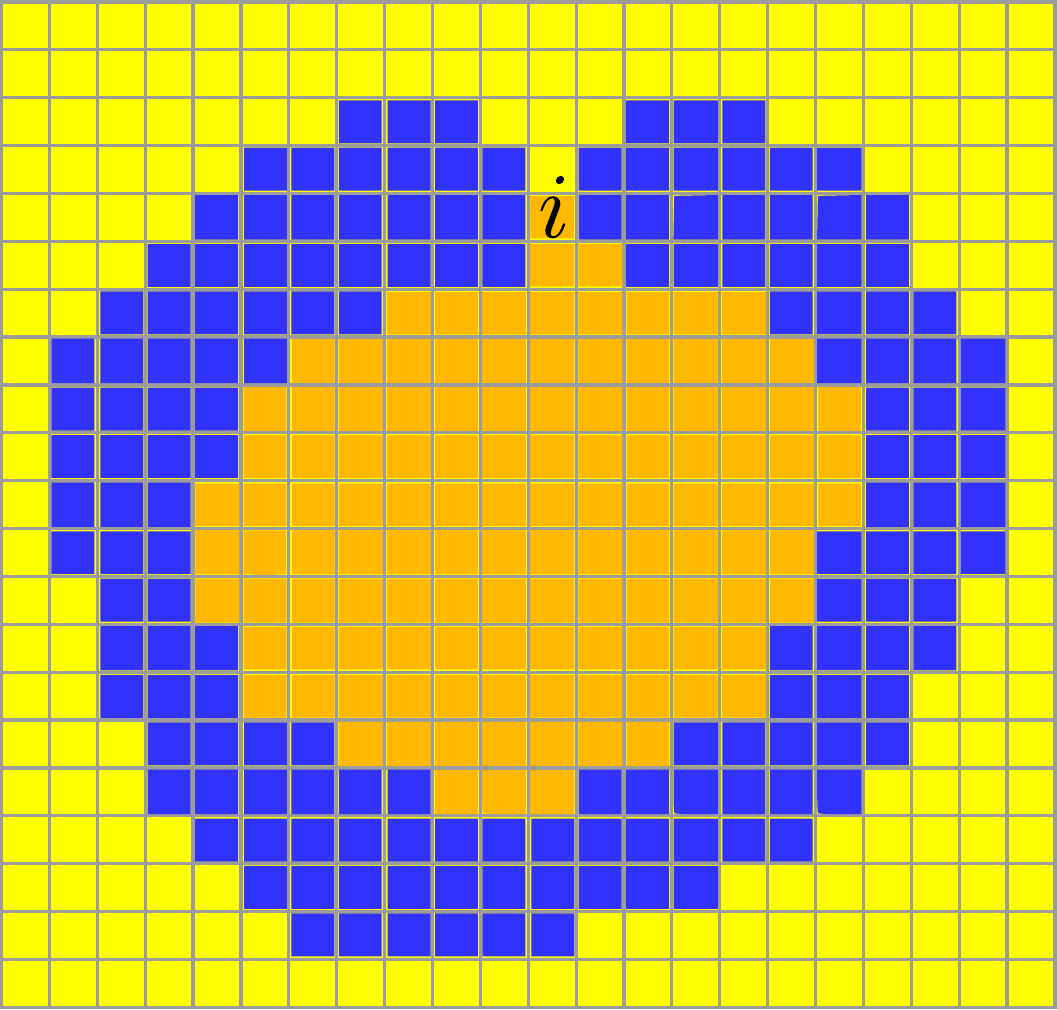}}
\caption{In a non-simply connected cell, the local connectivity of the candidate cell is not a necessary condition for the cell to stay connected after modification of site $i$ value: assuming that adjacency neigborhood $\Na$ coincides with Von Neumann or Moore neighborhood, the blue cell is not locally connected at \revision{site} $i$ in configuration (a) (nor in configuration (b)).
Yet, the cell stays connected 
when the site value is modified from configuration (a) to configuration (b). 
}
\label{connexity}
\end{figure}

It is also a necessary condition as long as the cell remains \textit{simply connected}, that is, homeomorphic to a disk. 
When the candidate cell is non-simply (\textit{aka} multiply) connected, our test is too restrictive, meaning that it is sufficient, but not necessary, to prevent fragmentation: the candidate cell can be non locally connected at a given site, although it remains connected after modification of this site value, as this is illustrated in Fig. (\ref{connexity}): 
the change of site $i$ from configuration (a) to configuration (b) is not allowed by our test, because the blue cell is not locally connected at site $i$. Yet, this modification would not fragment the cell.
Note also that in this situation, the detailed balance criterion is not satisfied anymore: in the example of Fig. \ref{connexity}, the reciprocal change from configuration (b) to configuration (a) is still allowed.

Since cell fragmentation is now forbidden by the local connectivity test, a cell must surround one (or more) other cell(s) to become multiply connected. When simulating foams, such an event is unlikely: the boundary energy per unit contact length $J_{\tau,\tau^\prime}$ is uniform, and it would require a very high temperature and very high size ratio in order for a larger bubble to surround a smaller one. In practice such a size ratio is hardly achievable, because we are limited by the finite size of the lattice.
When simulating biological cells on the other hand, such situation may occur when adhesion energies have high contrast. 

Multiple connected cells (and bubbles) are rather unrealistic. Fortunately, for a little extra computational cost, we can ensure that cells remain simply connected at all times.  
Remember that with our definition of local connectivity, a cell $\mathcal{C}$ can be locally connected at site $i$ even if $i$ is not in $\mathcal{C}$.  
If the target cell --- which is connected by construction (by choosing $\Nt \subseteq \Na$) --- is locally connected at site $i$, the modification of its value results in the formation of a ``handle'', and the target cell becomes multiply connected. 
Thus, to ensure that cells remain simply connected, we just have to additionally check that the target cell is locally connected at site $i$ before accepting the modification of the site value. In the example of Fig. \ref{connexity}, the change of site $i$ from configuration (a) to configuration (b) and the reciprocal change are now both forbidden.

In summary, the full CA consists of the following steps:
\begin{enumerate}[label=\arabic*c.]
\item Randomly select a lattice site $i$. Call this site the \textit{candidate site}. Let $\sigma$ be its value.
\item Randomly select a value $\sigma^\prime$ from those present in the target neighborhood $\Nt$. Call this the \textit{target value}.
\item Check the local connectivity of the candidate cell at site $i$, within the domain $\D_c$. If it is locally connected, proceed to next step. Otherwise, go to step 7c.
\item Check the local connectivity of the target cell at site $i$, within the domain $\D_c$. If it is locally connected, proceed to next step. Otherwise, go to step 7c.
\item Calculate the change in energy $\Delta E$ resulting in changing the lattice site value from $\sigma$ to $\sigma^\prime$.
\item Accept the site value modification with probability $\mathcal{A}(\sigma \rightarrow \sigma^\prime)=\min(1,e^{- \Delta E/T})$ (or any other acceptance probability that satisfies detailed balance equation).
\item Increment the number of copy attempts and go back to step 1c.
\end{enumerate}
Incidentally, target value draw has been slightly modified: its value is chosen arbitrarily in the set of different values present in the target neighborhood $\Nt$ (step (2c.)), without weighting by the number of neighboring sites that have this particular value, as it is in step (2b.) (see the Appendix for more details).
This modification, together with the two local connectivity tests that keep cells simply connected, restore detailed balance condition. 

The two flaws of the MMA are now solved: cell fragmentation is prohibited, and detailed balance condition is satisfied at \textit{all} temperatures.

Variants of this algorithm consist of swapping of steps (2c.) and (3c.), or (3c.) and (4c.). Relative efficiency of these variants depends on the simulation temperature $T$ and typical cell size $A_0$. To optimize the algorithm, local connectivity test with higher rejection rate should be tested first.
The probability that the candidate cell is locally connected at a given site decreases as $T$ increases. The probability that the target cell is not locally connected decreases as $T$ increases. The target cell becomes not locally connected at site $i$ when two cell protrusions come sufficiently close, what is unlikely at moderate simulation temperature. That is why we test it in last in our algorithm, but it could be more efficient to test it first at high temperature.

\revision{
Note that the number of copy attempts is incremented regardless of the results of the two connectivity tests (3c, 4c), in order to keep accurate correspondence between number of copy attempts and real time. Such correspondence is required for instance to satisfy equivalence of time and ensemble averages for systems at thermal equilibrium. 
}









\section{Choices for neighborhoods and local connectivity domain}\label{section_optimization}

\begin{figure}[htb]
\centering
\subfigure[\ ]{
\includegraphics[width=0.22\columnwidth]{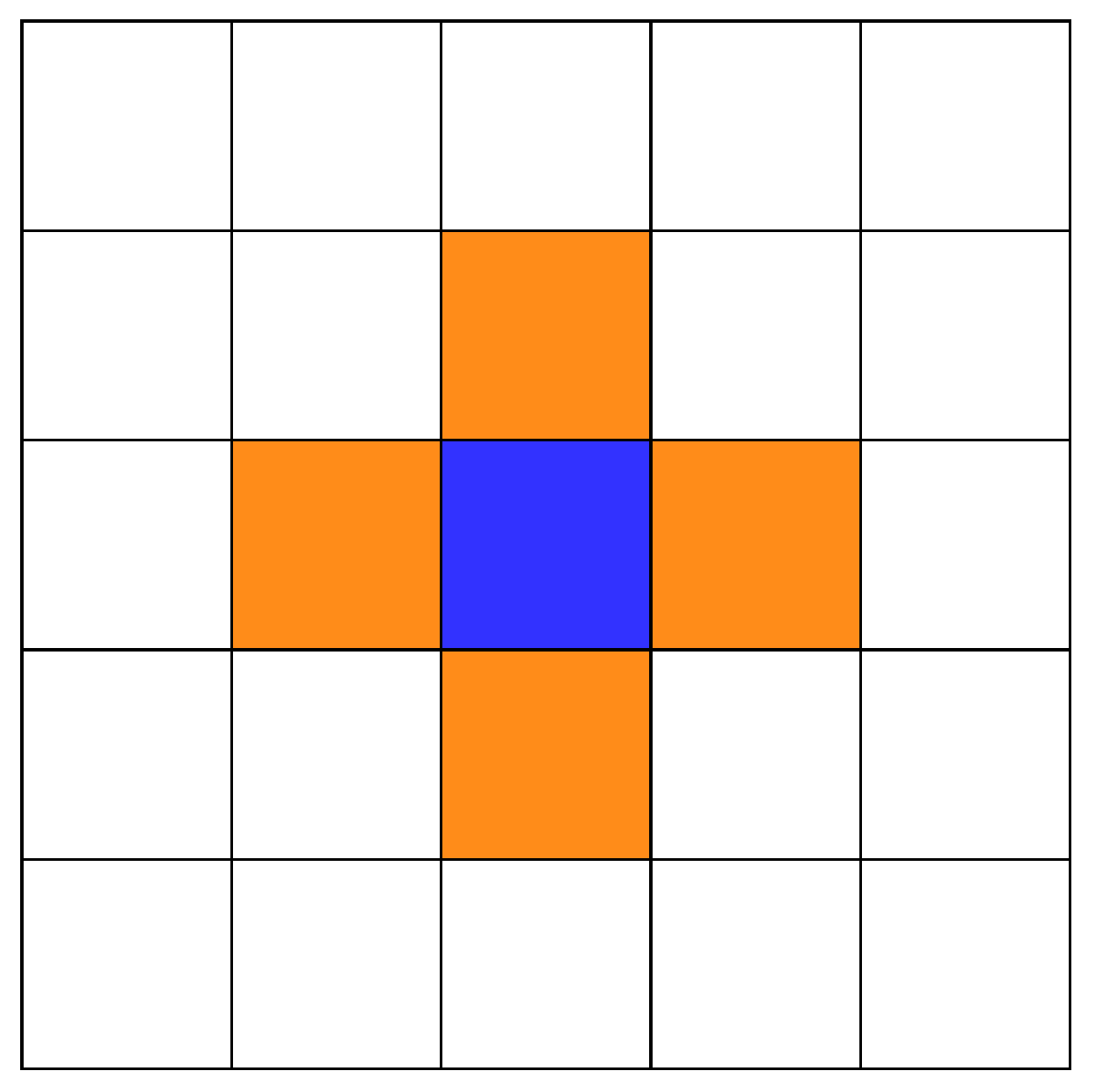}}
\subfigure[\ ]{
\includegraphics[width=0.22\columnwidth]{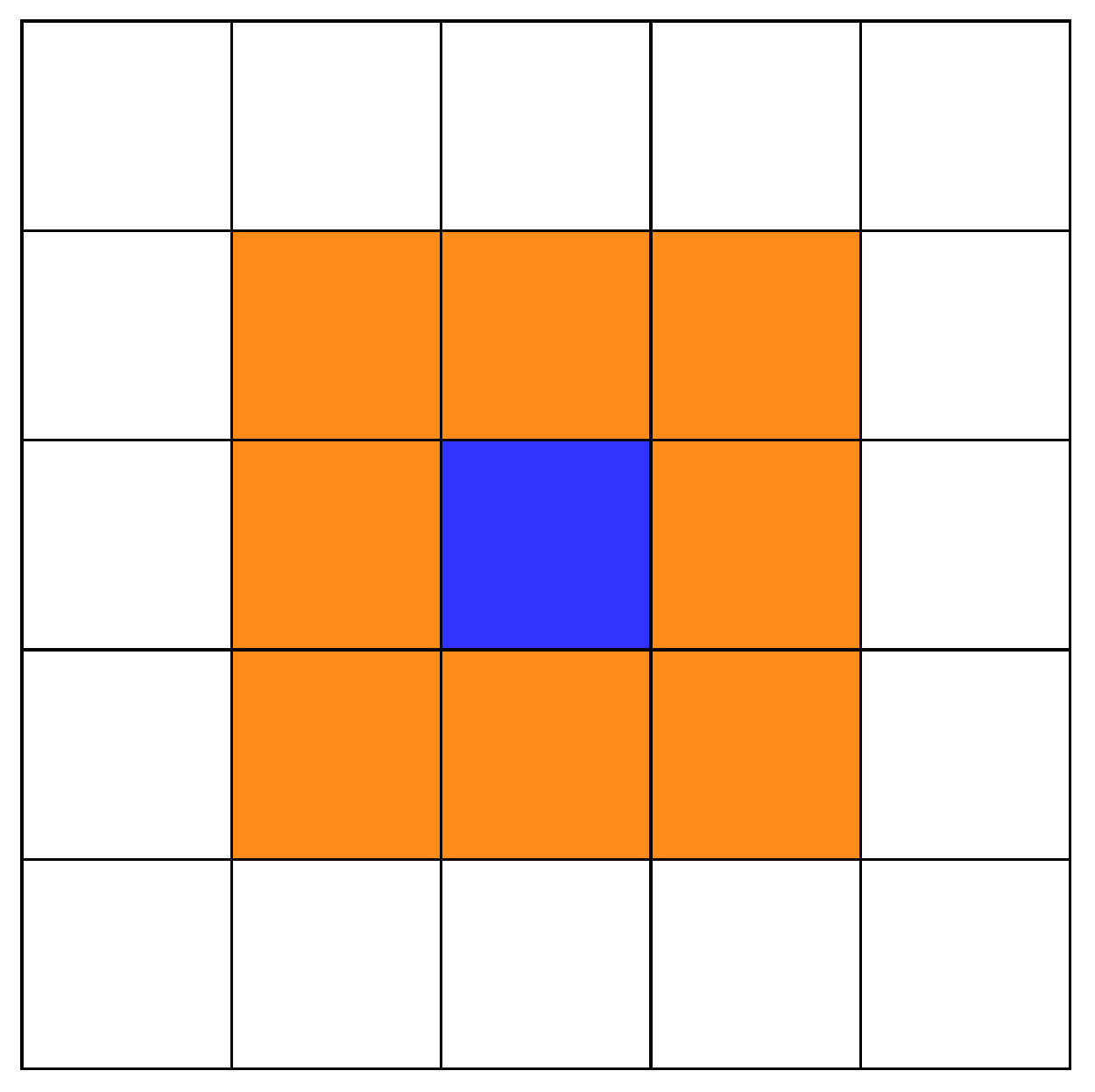}}
\subfigure[\ ]{
\includegraphics[width=0.22\columnwidth]{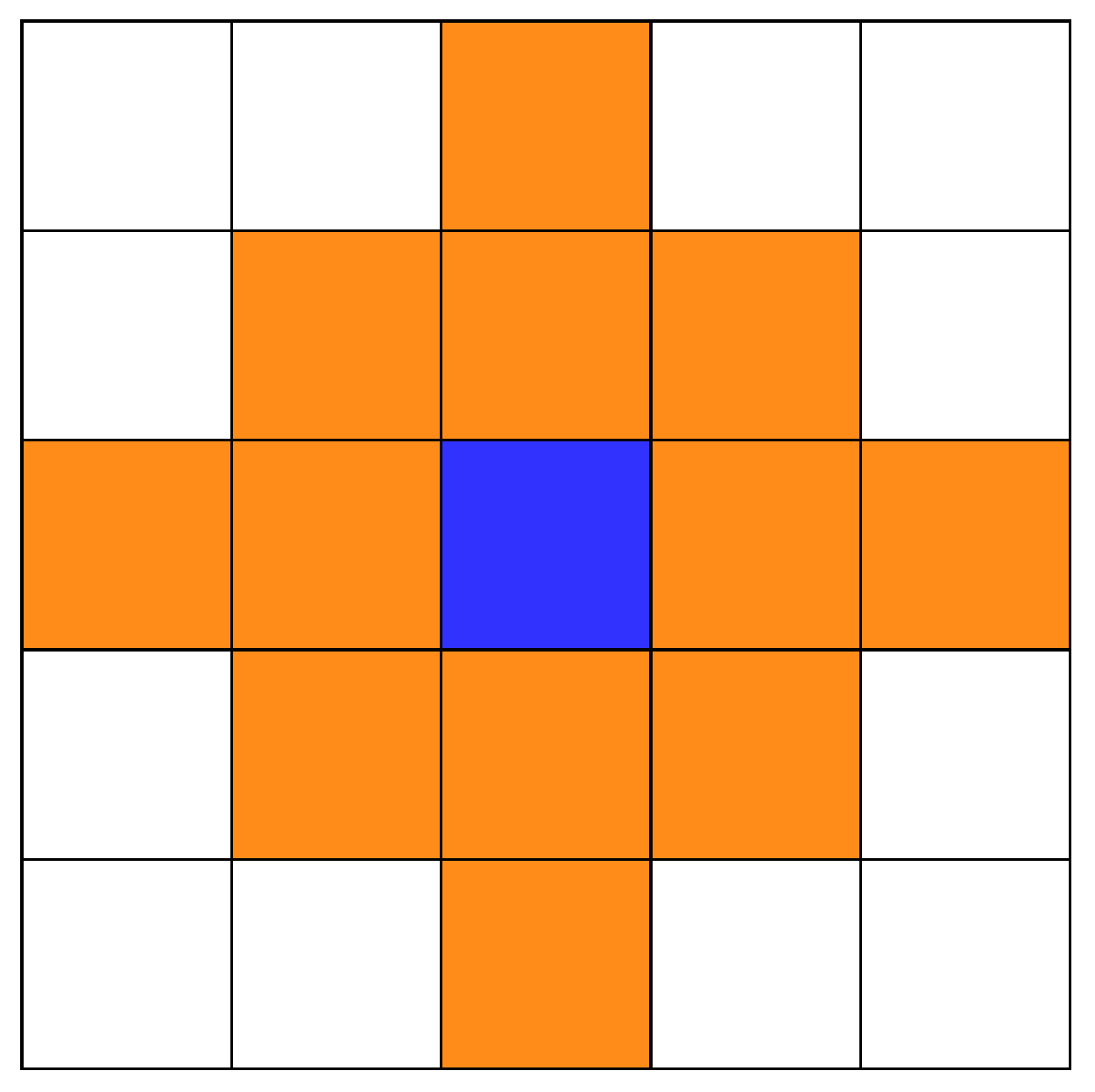}}
\subfigure[\ ]{
\includegraphics[width=0.22\columnwidth]{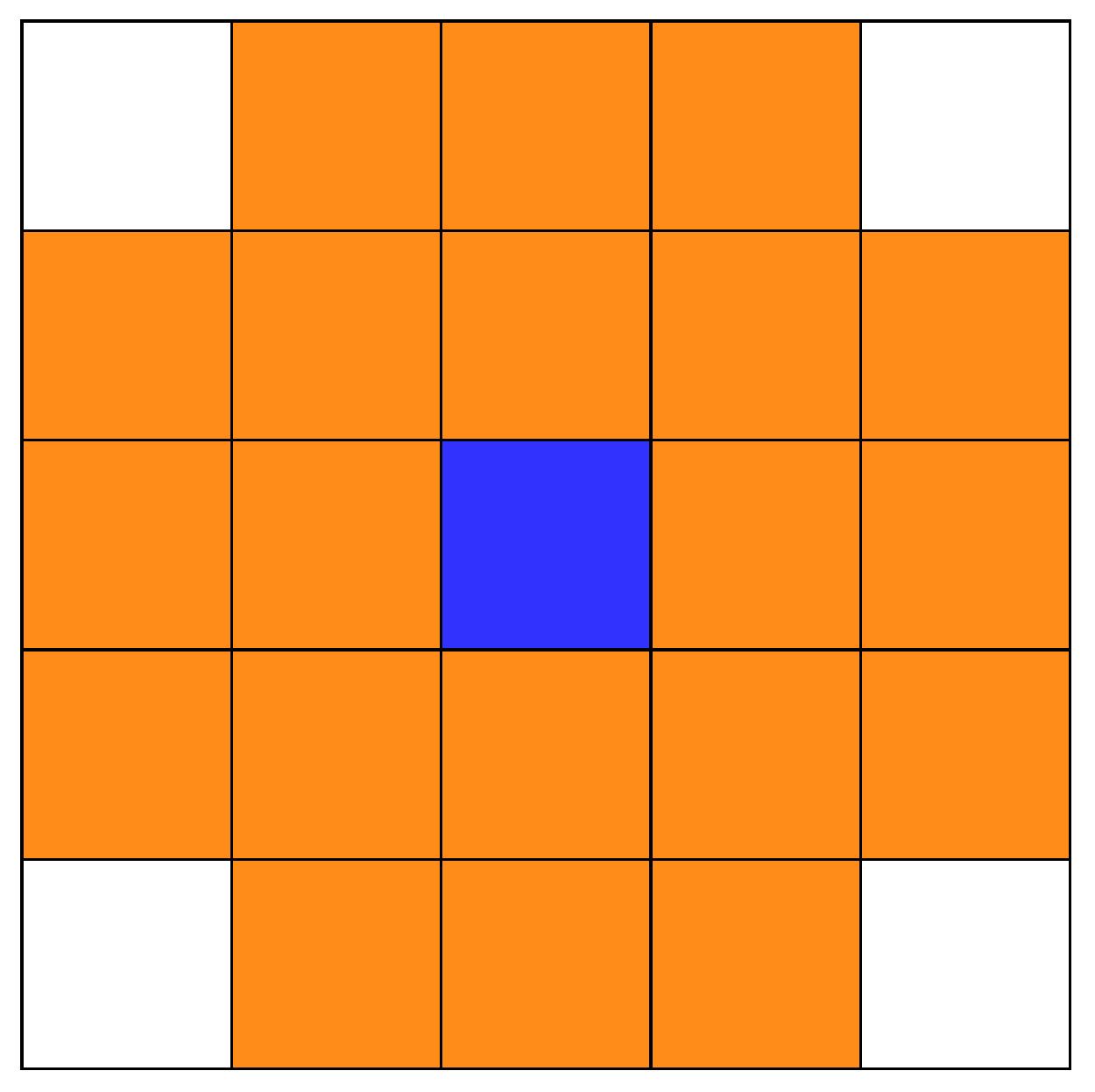}}
\caption{Different choices for the the neighborhood (in orange) of the central site (in blue),  in the square lattice: (a) order I (or Von Neumann) neighborhood; (b) order II (or Moore) neighborhood; (c) order III neighborhood; (d) order IV neighborhood. For adjacency and target neighborhoods, $\Na$ and $\Nt$ respectively, we choose order I neighborhood. For the local connectivity domain $\D_c$, we choose the order II neighborhood. For the coupling neighborhood $\Nc$, we use either order II or order IV neighborhood.
}
\label{neighbor-orders}
\end{figure}

Coupling neighborhood $\Nc$ is used in the calculation of the boundary energy, but does not play any role in the local connectivity test.
Taking a large domain for $\Nc$ reduces the lattice anisotropy, but increases the number and range of interactions, thereby slightly increasing simulation time. The range of interactions must also remain small compared with the typical cell size. Neighborhoods made of the 8 first neighboring sites (order II, or Moore neighborhood, Fig. \ref{neighbor-orders}b) and $20$ first neighboring sites (order IV neighborhood, Fig. \ref{neighbor-orders}d) are commonly used. 

Choices for the lattice sites that define the neighborhoods $\Na$ and $\Nt$ and local connectivity domain $\D_c$ strongly affect the efficiency of the local connectivity test, and must be chosen carefully.
%
Adjacency neighborhood $\Na$ is rarely defined in the literature, suggesting that it is often equated with the coupling neighborhood $\Nc$.
Yet, it would be better to choose it as small as possible to avoid overestimating the number of sides of cells. We then choose it equal to the order I -- or Von Neumann -- neighborhood, which is made of the $4$ lattice sites adjacent by side \revision{(Fig. \ref{neighbor-orders}a))}. 
This choice also avoids the unrealistic chessboard interlacing of cells that is observed at high temperature when choosing a larger adjacency neighborhood. More generally, two paths belonging to two distinct cells cannot cross each other. Hence, the definition of simple connectivity introduced in Sect. \ref{section_NMA} is a direct transposition to lattices of the definition \revision{that is} used for continuous topological spaces\footnote{Another definition for the simple connectivity of a domain $\mathcal{E}$, valid for any adjacency neighborhood, but restricted to 2D lattices, is as follows:  $\mathcal{E}$ is simply connected \textit{iff} it is connected and the complement of $\mathcal{E}$ is connected too.}.

The target neighborhood used in CPM simulations often corresponds with order II (or Moore) neighborhood. A more judicious choice is to take it identical to the adjacency neighborhood: as emphasized in Sect. \ref{section_NMA}, taking the target neighborhood smaller or equal to the adjacency neighborhood ensures that the target cell preserves its connectivity after each modification of site value. Thus, \revision{we recommend to take both equal to the order I (Von Neumann) neighborhood (Fig. \ref{neighbor-orders}a). }

For the local connectivity domain $\D_c(i)$, we choose the smallest domain as possible that contains the lattice sites in the adjacency neighborhood, plus the shortest paths that may connect them. 
\revision{With $\Na$ defined over the Von Neumann neighborhood,} $\D_c(i)$ then corresponds to the Moore neighborhood.

\begin{figure}[htb]
\centering
\subfigure[\label{neighbor-case-a}]{
\includegraphics[width=0.4\columnwidth]{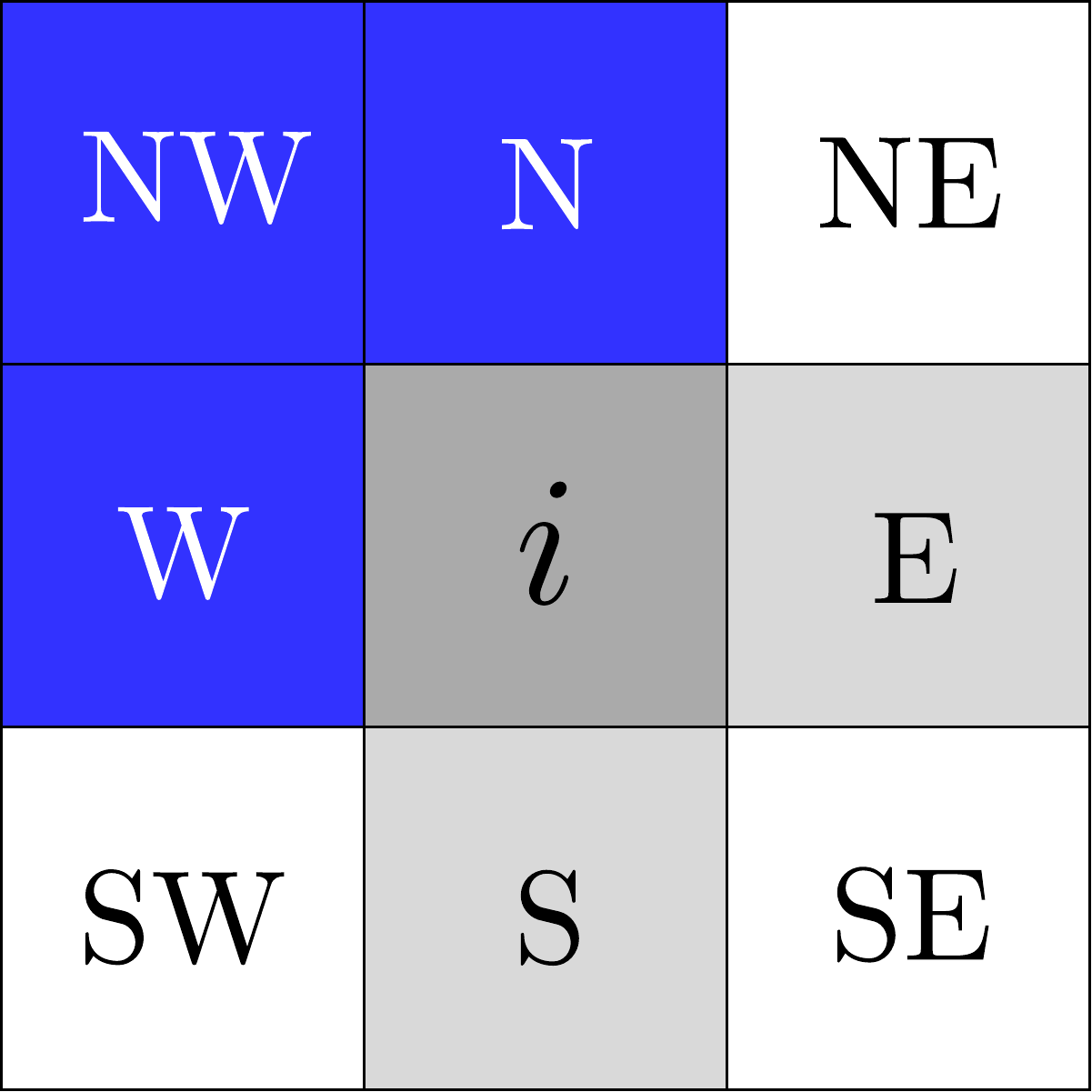}}
\hspace{0.1\columnwidth}
\subfigure[\label{neighbor-case-b}]{
\includegraphics[width=0.4\columnwidth]{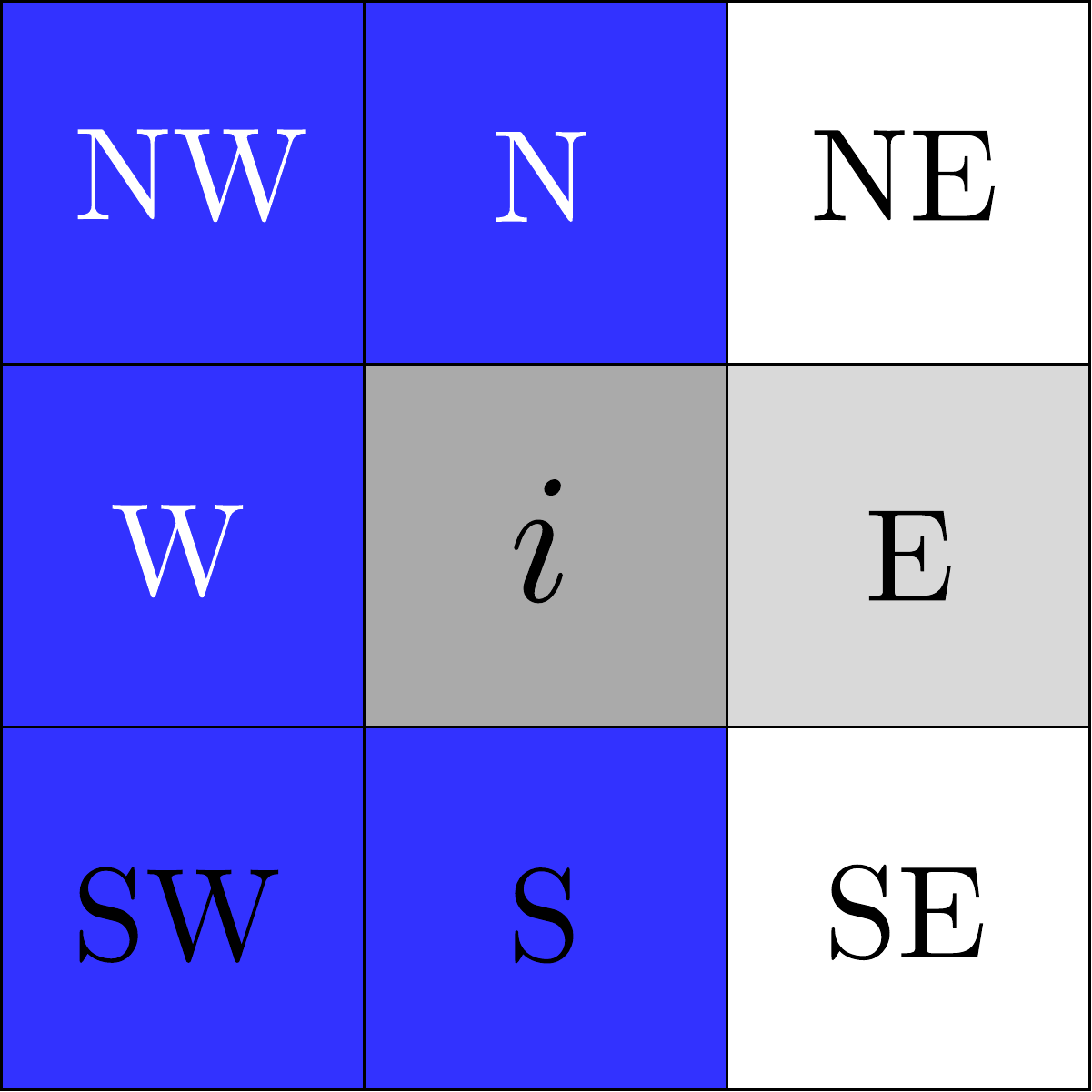}}
\caption{\label{neighbor-cases} Non-trivial patterns for which the blue cell is locally connected at site $i$, using an order I adjacency and target neighborhoods, and an order II local connectivity domain. (a) configurations with $z=2$ blue sites in the adjacency neighborhood of $i$: the two sites (here N(orth) and W(est)) must be corner-adjacent, and their common side-adjacent site (NW) must be blue too for the blue cell to be locally connected at $i$. (b) configurations with $z=3$ blue sites in the adjacency neighborhood of $i$: their two common side-adjacent sites must be blue too for the blue cell to be locally connected at $i$.}
\end{figure}

With this choice for $\Na$, $\Nt$ and $\D_c$, only a few patterns within the order II neighborhood need to be tested to check the local connectivity of a cell $\mathcal{C}$. Let $z$ be the number of lattice sites in the adjacency neighborhood of $i$ that belong to $\mathcal{C}$. Cases $z=0$, $z=1$, and $z=4$ are trivial: in the first case, $\mathcal{C}$ is never locally connected at site $i$. In the second case, $\mathcal{C}$ is always locally connected at site $i$. In the third case, value of site $i$ cannot be modified. When $z=2$, there are two possible situations: either the two neighbors face each other, or are corner-adjacent. Only the second situation can eventually lead to a locally connected cell, if their common side-adjacent site belongs to $\mathcal{C}$ too (see Figure \ref{neighbor-case-a}). When $z=3$, the two common side-adjacent sites must belong to $\mathcal{C}$ too for the cell to be locally connected (see Figure \ref{neighbor-case-b}). Accounting for the $\pi/2$ rotations of these admissible patterns, only four different patterns correspond to a locally connected cell for $z=2$, and similarly for $z=3$.

In practice, the implementation of the local connectivity test is as follows: we first detect, within the adjacency neighborhood of site $i$, the number ($z$) and positions (North, South, East, West) of the lattice sites that belong to cell $\mathcal{C}$ (either the candidate or target cell). If $z=2$ or $z=3$, we further check whether or not their positions and the values of their side-adjacent lattice sites match the patterns shown in Figs. \ref{neighbor-case-a} and \ref{neighbor-case-b} (modulo $\pi/2$ rotations), respectively.

\section{Benchmark}\label{section_benchmark}


To test the efficiency of the CA, we perform cell sorting simulations similar to those presented in the seminal work of Glazier and Graner \cite{graner_simulation_1992,glazier_simulation_1993}, and compare the results obtained with both CA and MMA. For simplicity, Hamiltonian (\ref{Hamiltonian}) has been used in the simulations presented here. To make a fair comparison, we used the same target, adjacency, and coupling neighborhoods and same connectivity domain for both algorithms. For the MMA, step (2b.) was replaced with step (2c.). The two algorithms then only differ by the test of local connectivity for the candidate and target cells in the CA. On a lattice with $N$ sites, we define one Monte Carlo Step (MCS) to be $N$ copy attempts. Simulations have been performed with and without $100$ annealing MCS. This unusually long annealing procedure allows us to compare the two algorithms on a large temperature range (for practical purpose, the program exits whenever the number of neighbors of a cell exceeds $20$). We duplicated the simulated pattern before annealing it to not alter the kinetics of cell sorting. 

Because of the two local connectivity tests within the CA, the correspondence between CPU time and MCS is different for the two algorithms: 1 MCS with CA takes a little bit more of CPU time than with the MMA. We checked that this difference is totally negligible: the increase of CPU time for similar runs ($6\times 10^6$ MCS) ranges from $0.235\%$ for low temperature ($T=35$) to $4.97\%$ for high temperature ($T=85$).

\begin{figure*}[htb]
    \subfigure[]{
        \includegraphics[width = 0.185\textwidth]{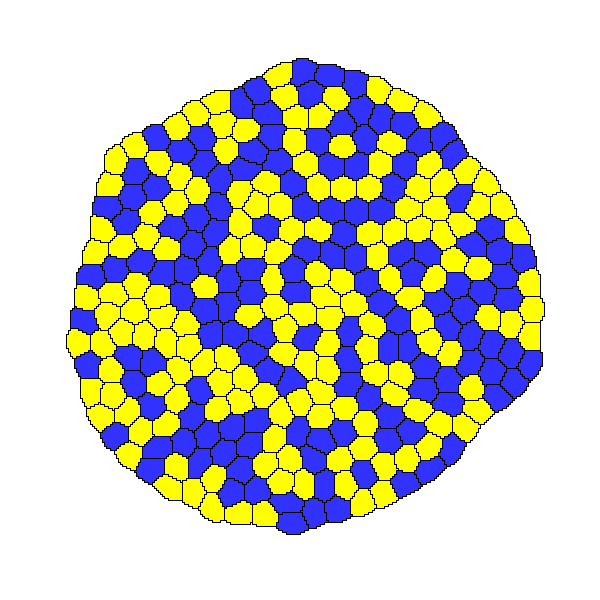}}
    \subfigure[]{
        \includegraphics[width = 0.185\textwidth]{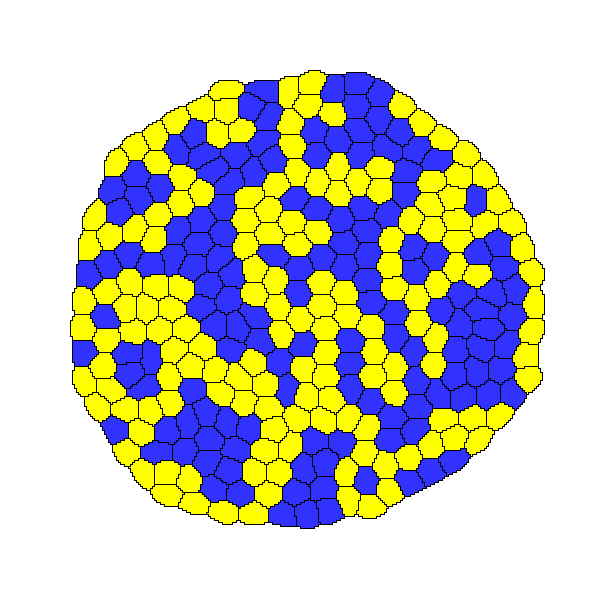}}
    \subfigure[]{
        \includegraphics[width = 0.185\textwidth]{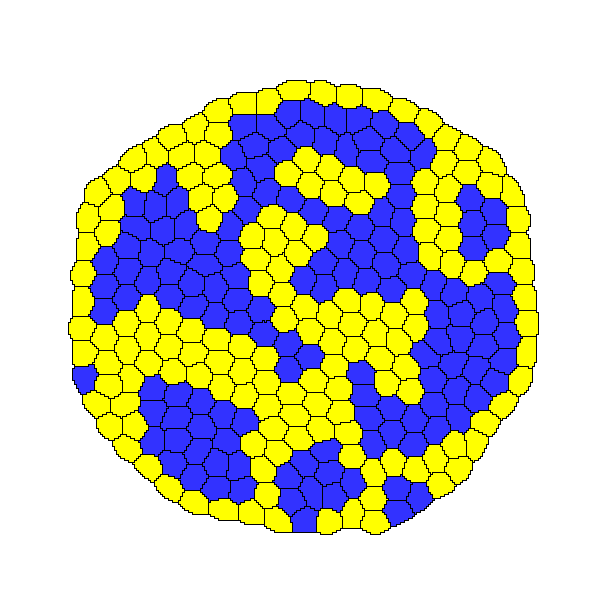}}
    \subfigure[]{
        \includegraphics[width = 0.185\textwidth]{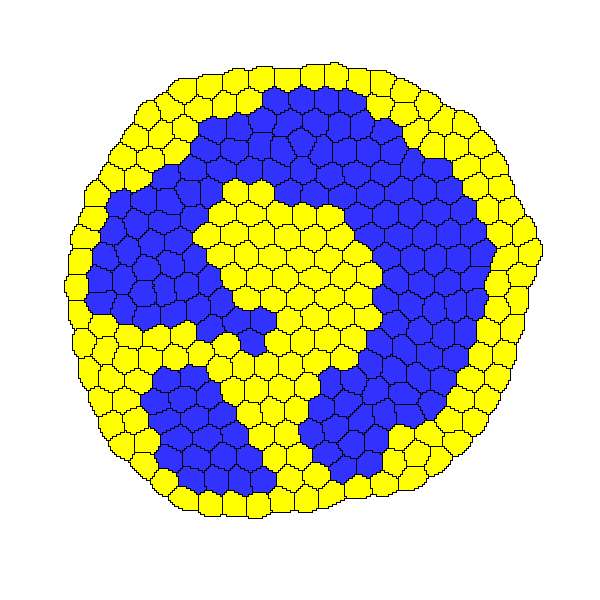}}
    \subfigure[]{
        \includegraphics[width = 0.185\textwidth]{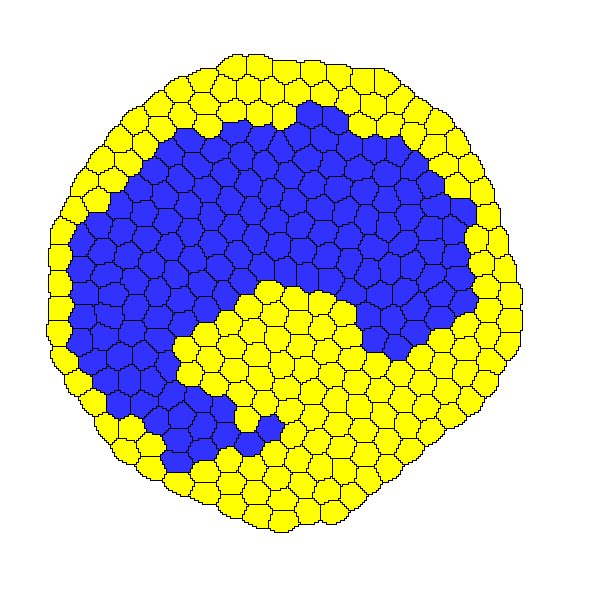}}
    \subfigure[]{
        \includegraphics[width = 0.185\textwidth]{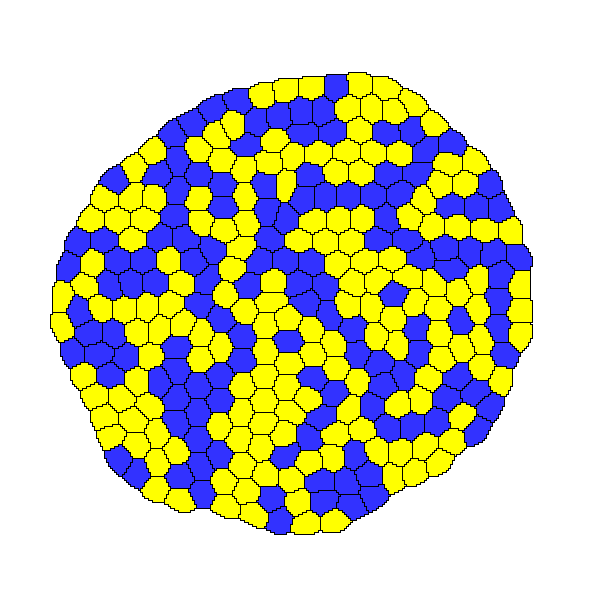}}
    \subfigure[]{
        \includegraphics[width = 0.185\textwidth]{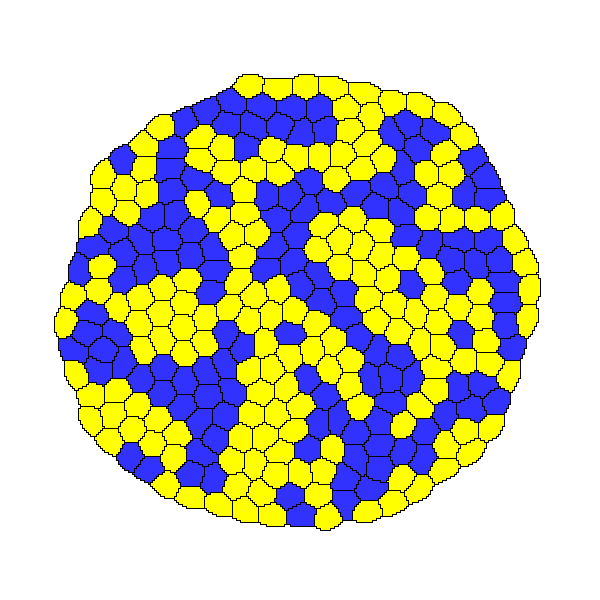}}
    \subfigure[]{
        \includegraphics[width = 0.185\textwidth]{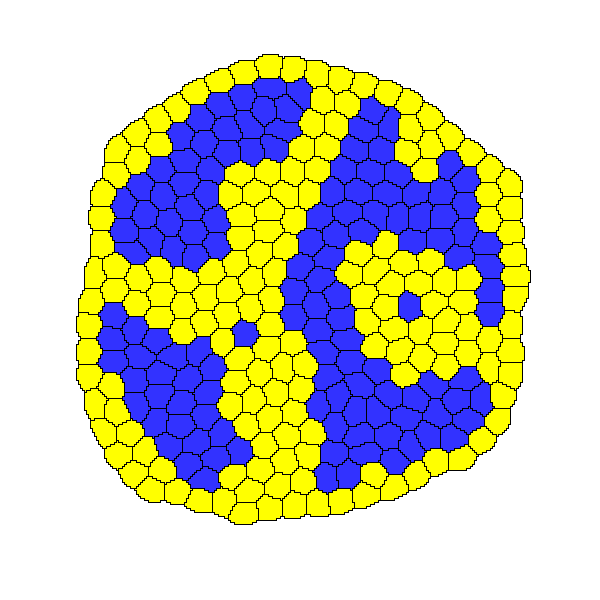}}
    \subfigure[]{
        \includegraphics[width = 0.185\textwidth]{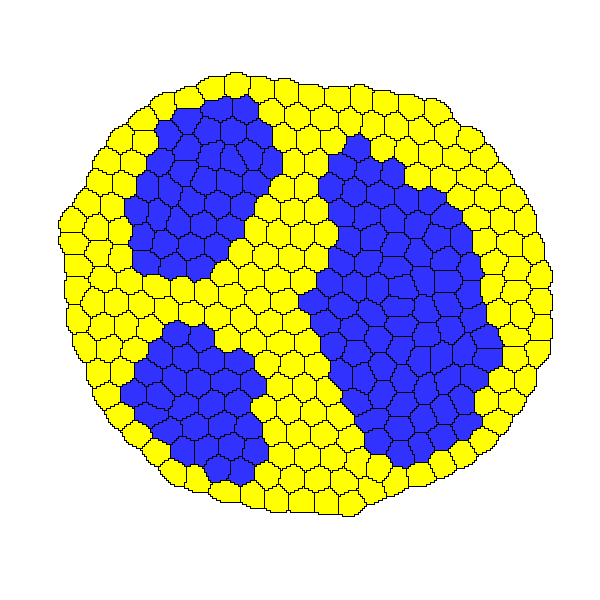}}
    \subfigure[]{
        \includegraphics[width = 0.185\textwidth]{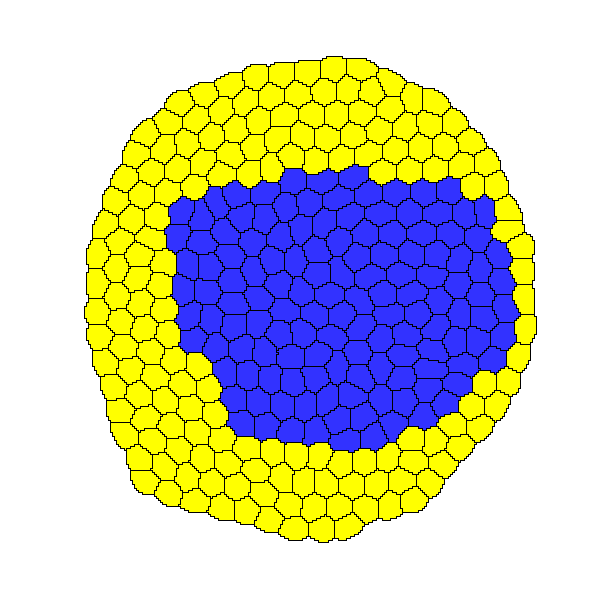}}
    \caption{Cell sorting simulations at low temperature ($T=35$). Patterns are displayed after 100 MCS of $T=3$ annealing steps. Top row: MMA. (a) $0$ MCS. (b) $10^4$ MCS. (c) $10^5$. (d) $10^6$ MCS. (e) $6\times10^6$ MCS. Bottom row: CA at same simulation times.}
    \label{fig:img_connexe_vs_non_connexe_low_T}
\end{figure*}

\begin{figure*}[htb]
    \subfigure[]{
        \includegraphics[width = 0.185\textwidth]{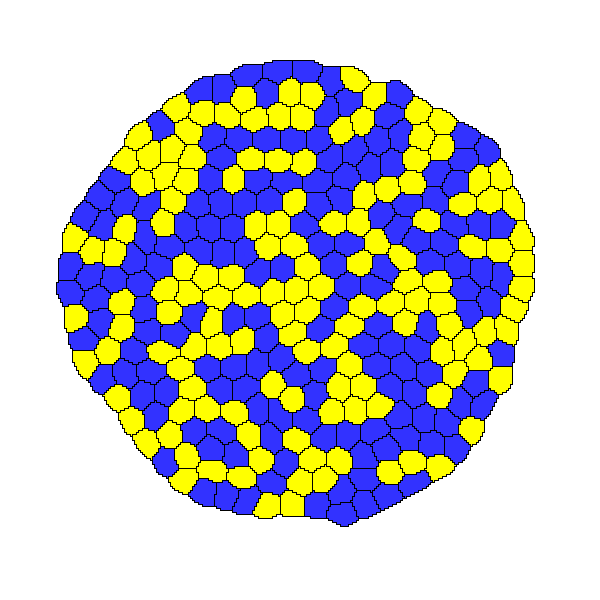}}
    \subfigure[]{
        \includegraphics[width = 0.185\textwidth]{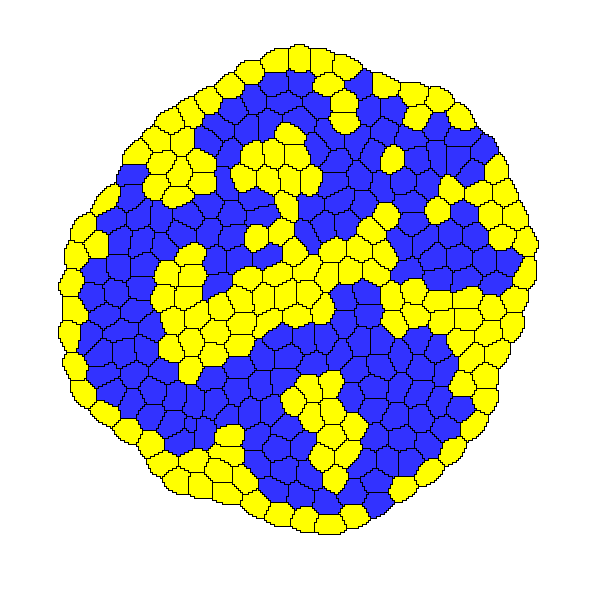}}
    \subfigure[]{
        \includegraphics[width = 0.185\textwidth]{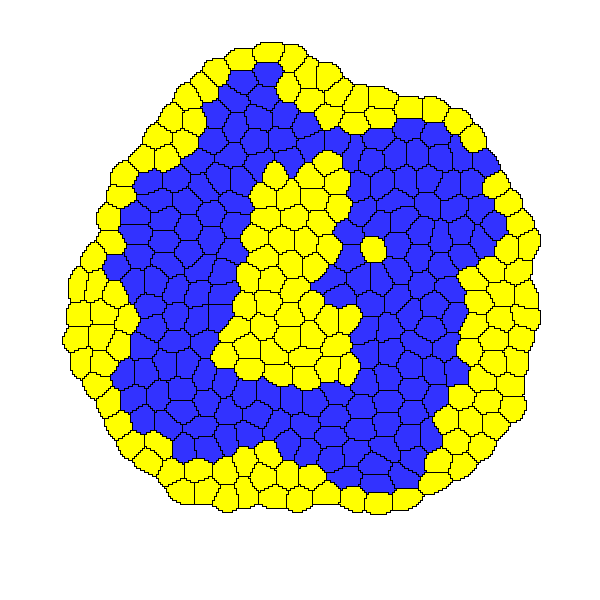}}
    \subfigure[\label{image_d}]{
        \includegraphics[width = 0.185\textwidth]{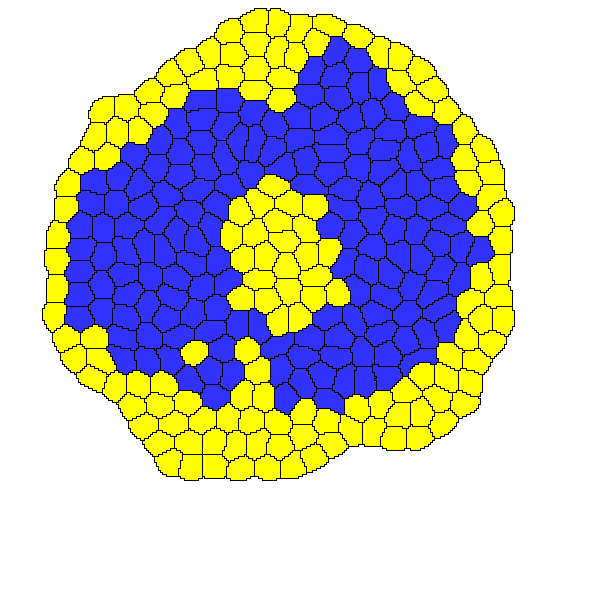}}
    \subfigure[\label{image_e}]{
        \includegraphics[width = 0.185\textwidth]{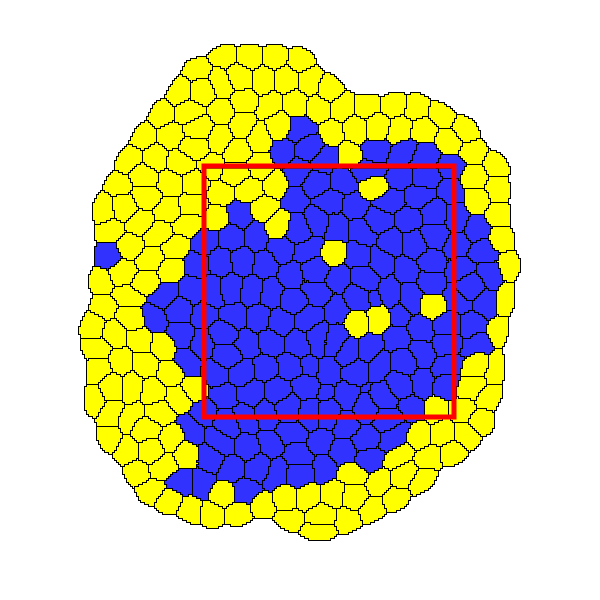}}
    \subfigure[]{
        \includegraphics[width = 0.185\textwidth]{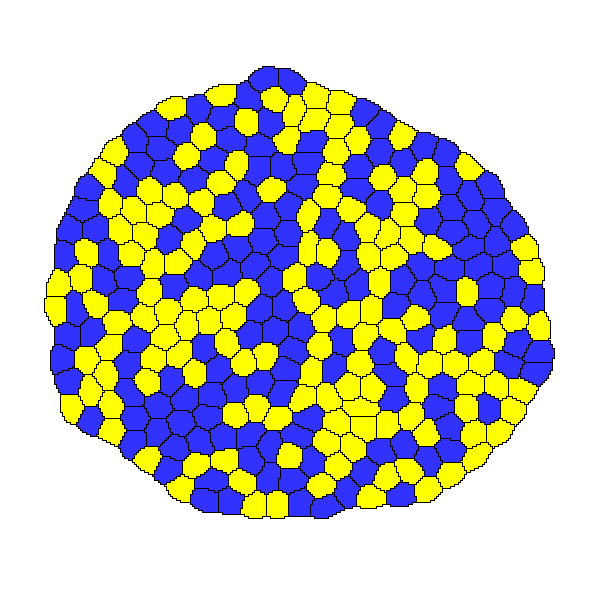}}
    \subfigure[]{
        \includegraphics[width = 0.185\textwidth]{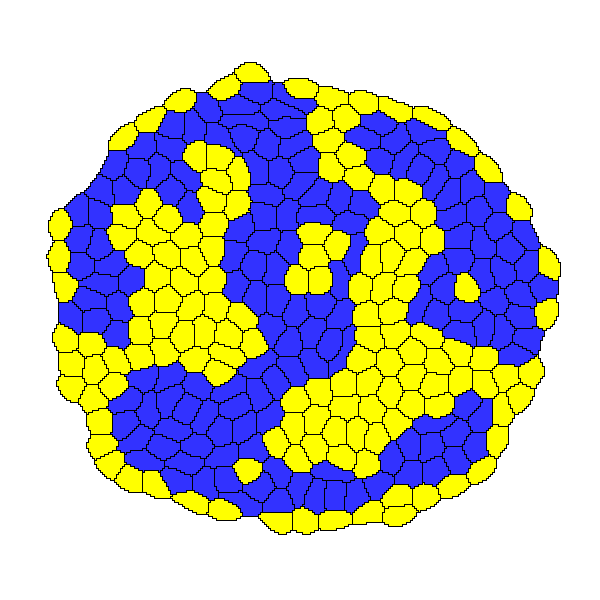}}
    \subfigure[]{
        \includegraphics[width = 0.185\textwidth]{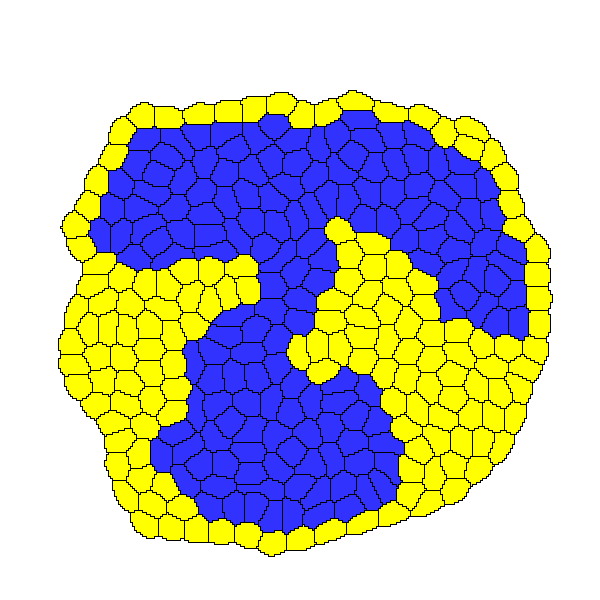}}
    \subfigure[\label{image_i}]{
        \includegraphics[width = 0.185\textwidth]{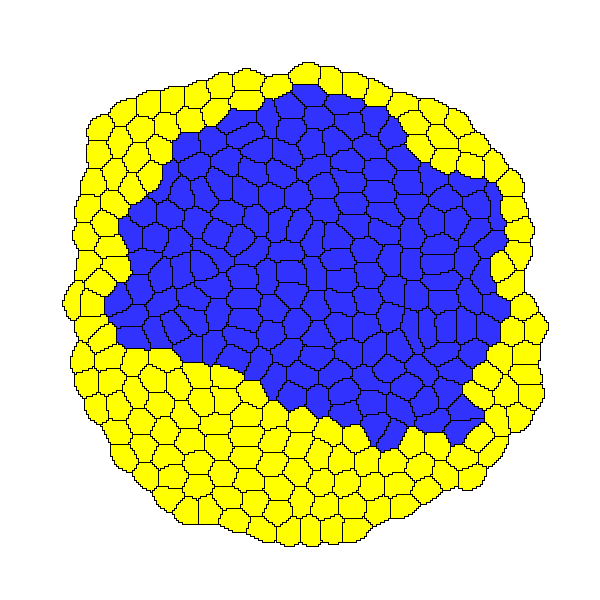}}
    \subfigure[\label{image_j}]{
        \includegraphics[width = 0.185\textwidth]{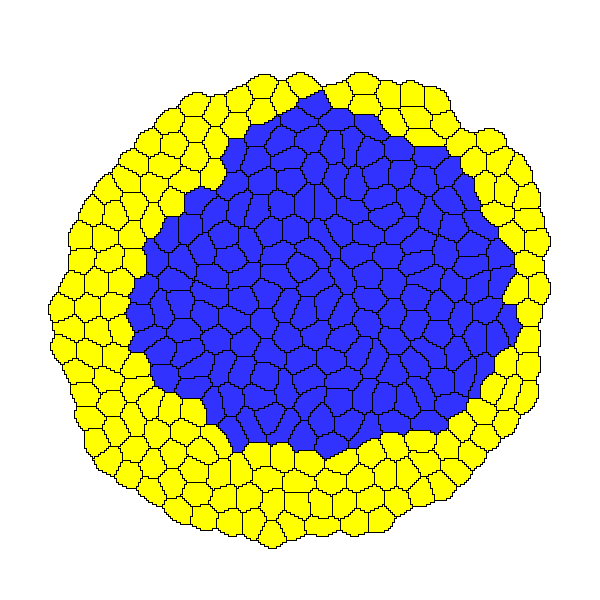}}
    \caption{Same as in Fig. \ref{fig:img_connexe_vs_non_connexe_low_T}, at high temperature ($T=85$). \revision{Red box indicates the area shown in Fig. \ref{fig:close-ups}(b).} }
    \label{fig:img_connexe_vs_non_connexe_high_T}
\end{figure*}

\begin{figure*}[htb]
    \subfigure[]{
        \includegraphics[width = 0.231\textwidth]{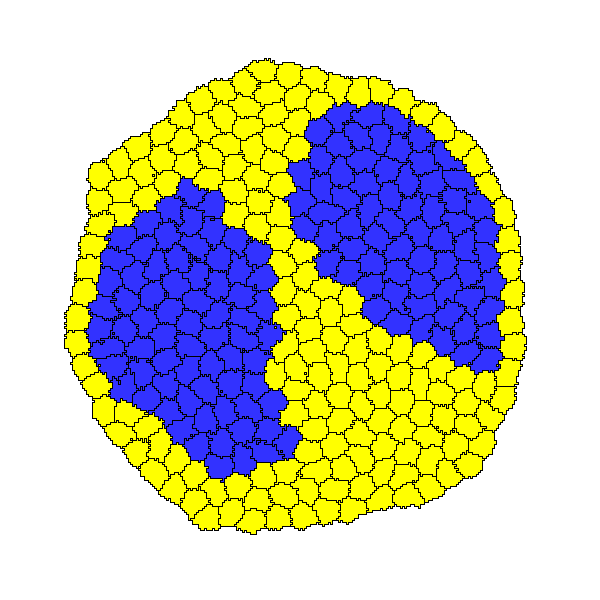}}
    \subfigure[]{
        \includegraphics[width = 0.231\textwidth]{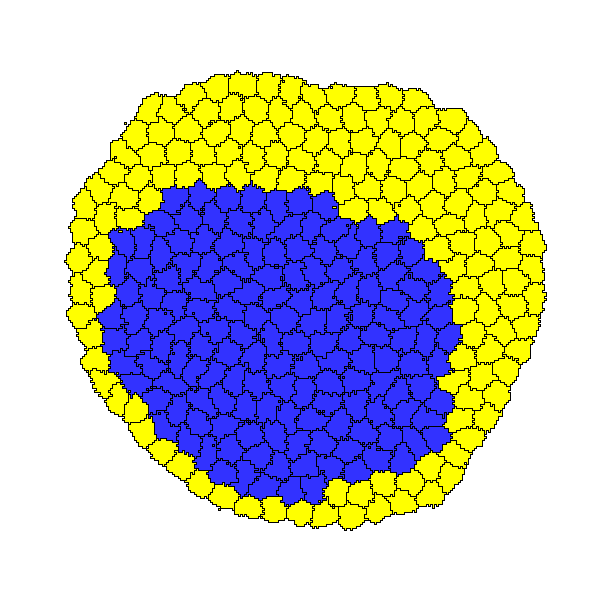}}
    \subfigure[]{
        \includegraphics[width = 0.231\textwidth]{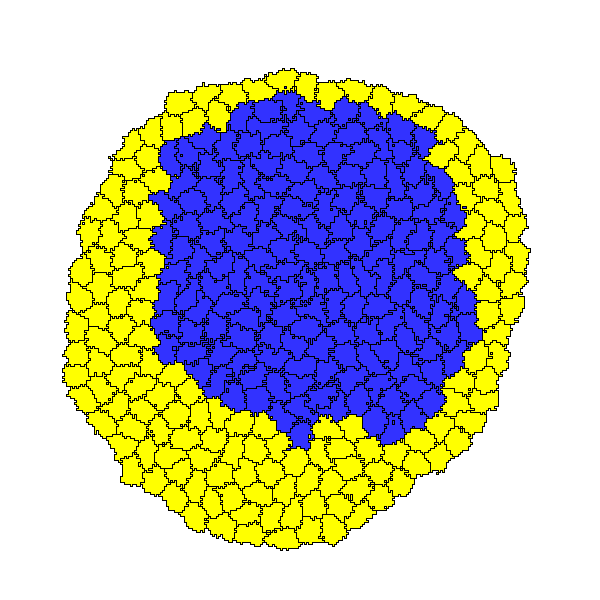}}
    \subfigure[\label{MMA_no_anneal}]{
        \includegraphics[width = 0.231\textwidth]{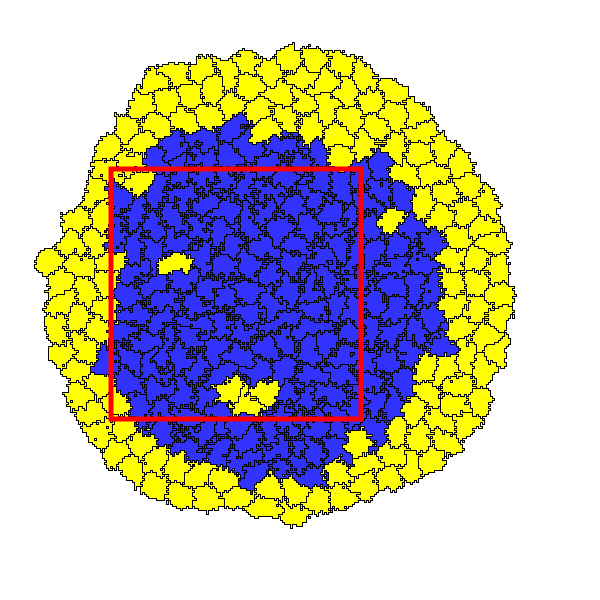}}\\
    \subfigure[]{
        \includegraphics[width = 0.231\textwidth]{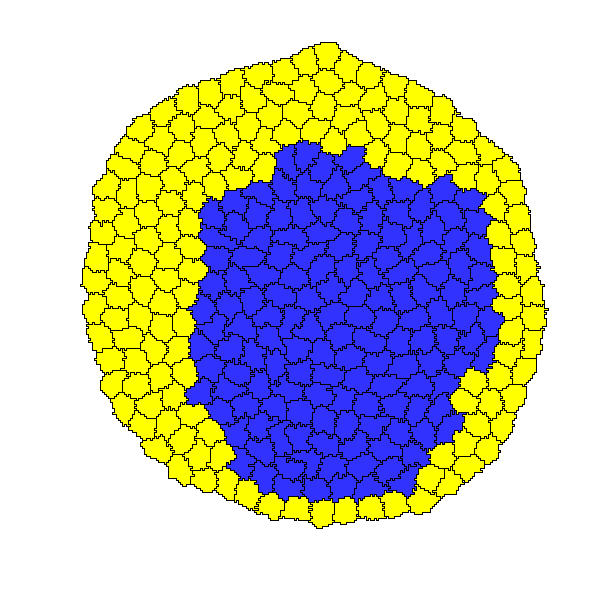}}
    \subfigure[]{
        \includegraphics[width = 0.231\textwidth]{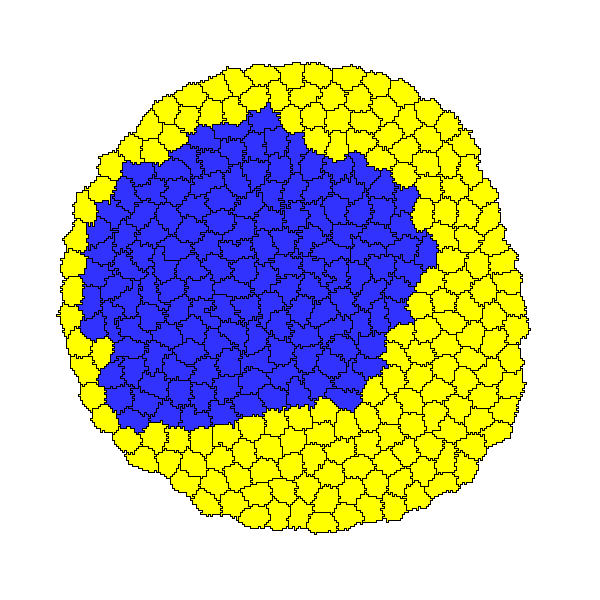}}
    \subfigure[]{
        \includegraphics[width = 0.231\textwidth]{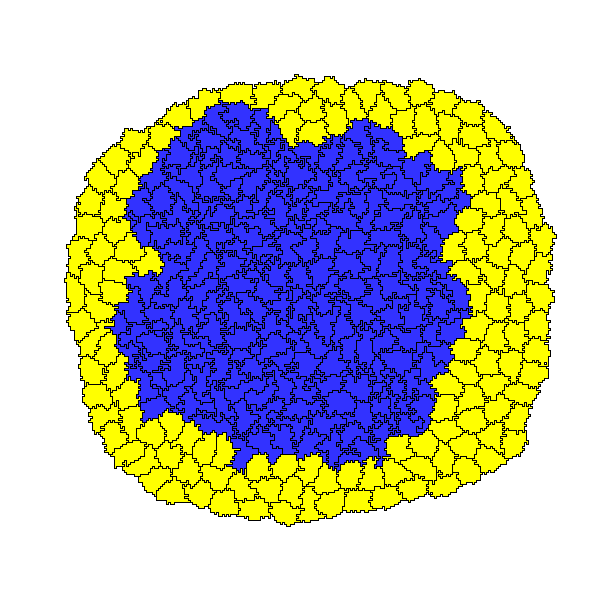}}
    \subfigure[]{
        \includegraphics[width = 0.231\textwidth]{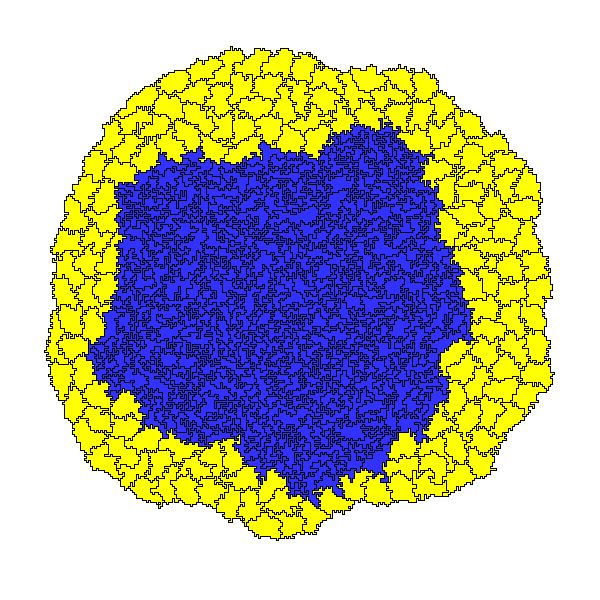}}
    \caption{Cell sorting patterns without annealing steps observed after $6\times10^6$ MCS. Top row: MMA. (a) $T=35$. (b) $T=40$. (c) $T=60$. (d) $T=85$.
    \revision{Red box indicates the area shown in Fig. \ref{fig:close-ups}(a).} Bottom row: CA at same temperatures.}
    \label{fig:without_anneal}
\end{figure*}

We start from an equilibrated rounded cluster of 150 B(lue) and 150 Y(ellow) cells, randomly positioned. The cluster is surrounded with the M(edium), and the total number of sites is $N=300\times 300$.
Values of the different parameters used in the Hamiltonian (\ref{Hamiltonian}) are: $J_{BM}=16$, $J_{YM}=16$, $J_{BB}=8$, $J_{YY}=14$, $J_{YB}=12$, $A_0=150$, $B=200$.
Thanks to our discrimination of the different neighborhoods, the size of coupling neighborhood has a very limited effect on the computing time.
We choose an order IV coupling neighborhood, which is sufficiently large to reduce the lattice anisotropy, whilst remaining small compared with the typical cell size. The other neighborhoods and local connectivity domain are chosen as recommended in Sect. \ref{section_optimization}.
Typical evolution of the annealed patterns are shown in Figs. \ref{fig:img_connexe_vs_non_connexe_low_T} and \ref{fig:img_connexe_vs_non_connexe_high_T}. We also show on Fig \ref{fig:without_anneal} typical unannealed patterns observed after $6\times 10^6$ MCS.

\begin{figure}[htb]
\centering
\subfigure[]{
 \includegraphics[width = 0.22\textwidth]{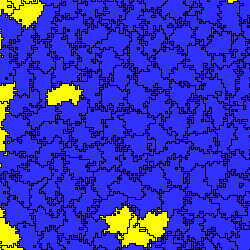}}
 \hfill
 \subfigure[]{
 \includegraphics[width = 0.22\textwidth]{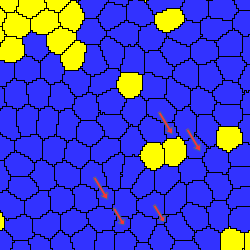}}
  \caption{Details of the patterns obtained with the MMA at $T=85$, after $6\times10^6$ MCS. (a) Without annealing (close-up of Fig. \ref{MMA_no_anneal}): the pattern exhibits numerous  fragments.  \revision{(b)} After 100 annealing steps (close-up of Fig. \ref{image_e}): the pattern still exhibits large fragments, indicated by arrows.}
\label{fig:close-ups}
 \end{figure}
 
We first compare the mean number of sides per cell $\langle n \rangle$ obtained with the two algorithms.
According to Euler theorem \cite{Durand_EPJE_2015}, $\langle n \rangle$ is equal to $6$ in a pattern containing only 3-fold vertices, if the surrounding medium is counted as an extra cell. Discrepancy with this theoretical value can have several origins: \textit{i)} due to the discretization of the pattern on the lattice, two 3-fold vertices can be mingled with one 4-fold vertex. This artifact can only decrease the measured value of $\langle n \rangle$, and is reduced when the target area $A_0$ (and so the typical edge length) is increased. \textit{ii)} The algorithms actually evaluate the number of different neighboring cells -- rather than the number of sides -- of each cell. These two numbers differ when two neighboring cells have two sides in common (thereby encircling a third cell). This would underestimate $\langle n \rangle$. In practice, configurations where two cells share more than one side are unlikely after 100 annealing steps, and the two numbers are equal.  \textit{iii)} when a cell is fragmented, its number of neighboring cells increases or stays constant. Thus, $\langle n \rangle$ cannot decrease under fragmentation.


In Table \ref{table} are reported the average and standard deviation values of the number of neighboring cells, with and without $100$ annealing steps. 
The annealing temperature is $T_{anneal}=3$. Data are averaged over $20$ runs.
In presence of annealing steps, values obtained with both algorithms are close to the theoretical value $6$. 
Notice that at high temperature ($T=85$) though, the averaged value is slightly above $6$. Indeed, Fig. \ref{fig:close-ups} reveals the presence of large fragments, despite the unusually long annealing procedure (only a few annealing steps are usually performed in literature). Without annealing, values obtained with the MMA are above $6$, and increase with $T$, due to cell fragmentation. Values obtained with the CA are equal to, or slightly below $6$, and slightly decrease as $T$ increases, due to the presence occasionally of 4-fold vertices.
Note that for the CA, the values reported with and without annealing steps, and for any temperature, are very close, which confirms that no fragmentation occurs.
%
\begin{table*}[ht]
\begin{tabularx}{\textwidth}{|c||Y|Y||Y|Y|}
\hline 
 & MMA \newline without annealing & MMA \newline after 100 annealing steps & CA \newline without annealing & CA \newline after 100 annealing steps \\ 
\hline 
$T=35$ & $6.001 \pm 0.005$ & $6.000 \pm 0.001$ & $5.995 \pm 0.005$ & $5.999 \pm 0.002$ \\ 
\hline 
$T=40$ & $6.004 \pm 0.008$ & $5.999 \pm 0.002$ & $5.992 \pm 0.007$ & $5.999 \pm 0.003$ \\ 
\hline 
$T=60$ & $6.049 \pm 0.020$ & $5.999 \pm 0.003$ & $5.982 \pm 0.011$ & $5.999 \pm 0.003$ \\ 
\hline 
$T=85$ & $6.276 \pm 0.047$ & $6.001 \pm 0.007$ & $5.980 \pm 0.011$ & $5.995 \pm 0.005$ \\ 
\hline 
\end{tabularx}
\caption{Mean and standard deviation values of the number of neighbours of a cell obtained with the standard CPM algorithm (MMA) and the new proposed algorithm (CA), with and without $100$ annealing steps \label{table}}
\end{table*}

\begin{figure*}[htb]
\centering
\subfigure[]{
    \includegraphics[height = 6.2cm]{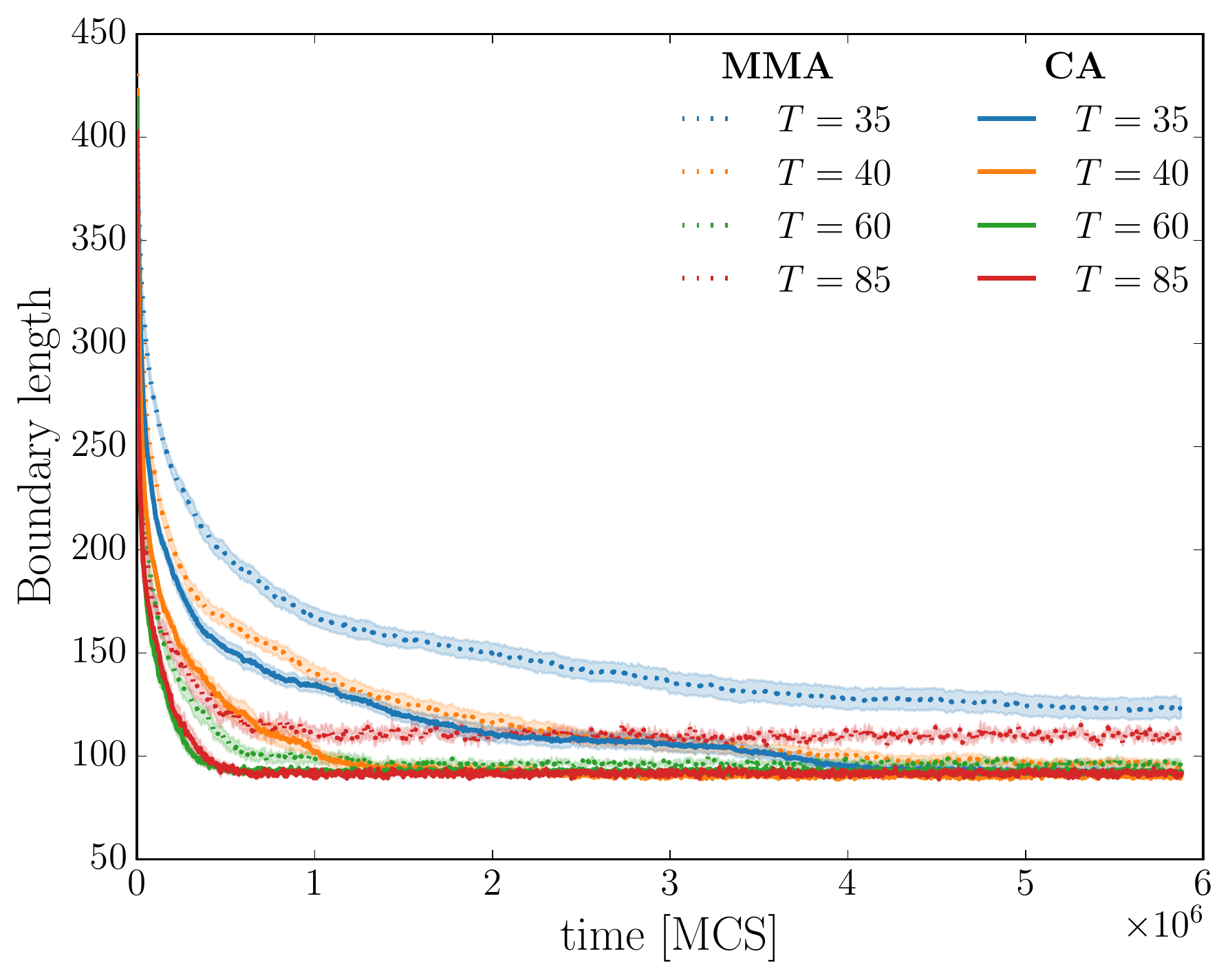}
    \label{benchmark}}
\hfill
\subfigure[]{
    \includegraphics[height = 6.2cm]{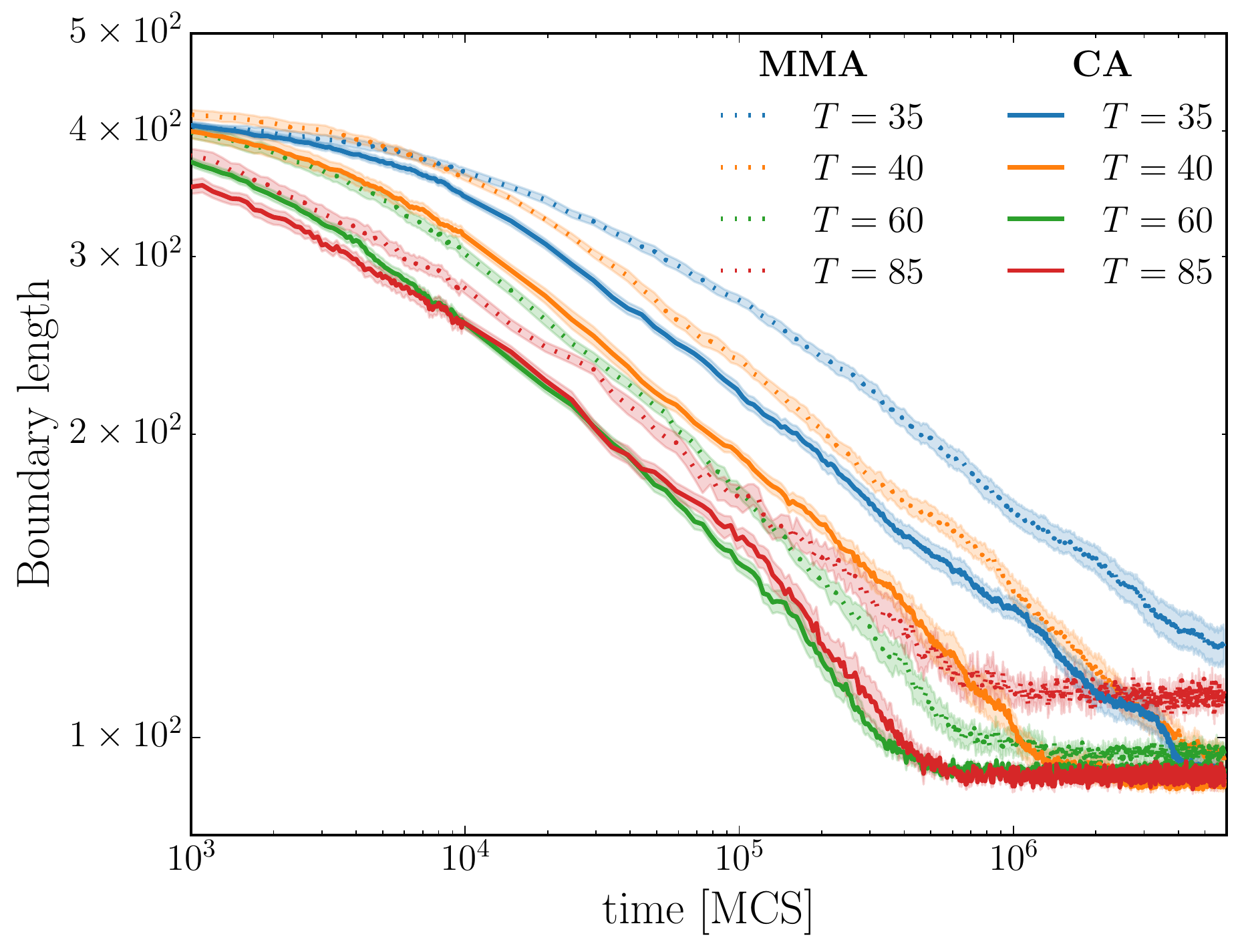}
    \label{benchmark-log-log}}
\caption{(a) Boundary length (number of edges that are shared by blue and yellow cells) as a function of time, for four different simulation temperatures: $T=35$, $T=40$, $T=60$, and $T=85$. Shaded areas mark standard errors of the mean. Values are measured after $100$ MCS of $T=3$ annealing steps, and averaged over $20$ runs; (b) same data in log-log scale.
}
\label{fig:interface}
\end{figure*}

Then, we compare the evolution of the \textit{boundary length}, defined as the number of edges that are shared by blue and yellow cells \cite{graner_simulation_1992,glazier_simulation_1993,Nakajima_2011}. Values \revision{shown} in Figure \ref{fig:interface} are those obtained after $100$ annealing steps.
The boundary length at initial time is close to the theoretical value $450$ for randomly positioned blue and yellow cells, then it decreases -- with a temperature-dependent rate -- down to a plateau referred to as \textit{long-term stage}.

At low temperature  ($T=35$), the CA converges to the long-term stage much more rapidly than the MMA. This difference highlights that the few fragmentation which occurs with the MMA between annealing steps has a dramatic effect on the dynamics of cell sorting.
It is not clear if complete cell sorting will eventually be reached with the MMA: as noticed before \cite{glazier_simulation_1993}, a minimal temperature seems required to achieve complete cell sorting, at least with the MMA.
Actually, we were not able to find a set of parameter values for which complete sorting is achieved with identical kinetics for both algorithms, which suggests that the temperature range in which fragmentation has a negligible effect in the cell sorting kinetics is above the critical temperature required to achieve complete sorting.

As temperature is slightly increased ($T=40$), both algorithms converge to the same stage: a complete cell sorting with round cluster of blue cells surrounded by yellow cells. But the convergence is still much faster with the CA.
The long-term value ($\simeq 90$) is consistent with a rough estimation in which the blue cells are assumed to form a circular cluster of radius $R=\sqrt{N_B}d/2$, where $N_B=150$ is the number of blue cells in the cluster, and $d=2\sqrt{A_0/\pi}$ is the diameter of the cells:
since all junctions are trivalent, the number of edges that belong to the boundary is equal to $N^-+N^+$, where $N^-\simeq 2\pi R/d$ and $N^+\simeq 2\pi (R+d)/d$ are the number of cells in the inner and outer shell of cells at the boundary. Hence, $N^-+N^+\simeq 2\pi (\sqrt{N_B}+1)\simeq 83$. This constitutes a lower bound, as non-circular clusters have a larger boundary length.


At moderate temperature ($T=60$), the CA is still more efficient, although the difference in time evolution is less pronounced: the increase of fragment sizes in MMA favorizes large (but unrealistic) displacements of the cell centroids. 

At high temperature ($T=85$), the behavior of both algorithms differs clearly: for the CA, the long time stage is the same as at low and moderate temperatures -- a complete cell sorting with blue cells surrounded by the yellow cells. Thanks to detailed balance criterion, once the thermal equilibrium is reached, the long-term stage is the same whatever the path we choose in the simulation temperature space: it depends only on the final temperature, which is here the annealing temperature.
For the MMA on the other hand, evolution of the boundary length is noisier and the plateau at long times is noticeably higher. Figure (\ref{image_e}) reveals that the blue cell cluster is highly contorted. Moreover, complete cell sorting is never attained: some yellow cells are constantly entering and leaving out the blue cell cluster, and some blue cells are ejected from it, what opposes to sorting process. 
This phenomenon, inherent to fragmentation (and violation of detailed balance), is never observed with the CA. 

Note that for both algorithms the time evolution of the boundary length is approximately the same than at moderate temperature. The gain in cell sorting process becomes marginal, as spatial spread and entanglement of the cells restrain their mobility. 

Log-log plots (Fig. \ref{benchmark-log-log}) show that cell sorting kinetics obtained at long times with both algorithms is compatible with a power law dependence, although our sample is too small to discriminate between logarithmic or power-law decay \cite{Nakajima_2011}. 

We performed simulations at higher temperatures, up to $T=150$ (not shown). With the CA, the kinetics of cell sorting is approximately the same, and $\langle n \rangle$ remains close to $6$, with or without annealing steps. With the MMA, the number of cells entering or leaving the blue cell cluster increases, and the value of the long-term plateau gets higher as the temperature increases. $\langle n \rangle$ also significantly increases, up to $6.46$, in spite of $100$ annealing steps. $\langle n \rangle$ cannot be evaluated without annealing, as the number of neighbours of a cell often exceeds the limit value $20$ set in the program, due to the high fragmentation rate.

\section{Conclusion and Outlook}\label{Outlook}

To summarize, we provide a new algorithm for CPM simulations that forbids cell fragmentation (in addition to spontaneous nucleation) by testing the local connectivity of the candidate and target cells before every modification of a site value. It is shown that these two local connectivity tests are rigorously equivalent to testing the simple connectivity of the cells. This algorithm presents numerous advantages (and no drawbacks have been identified):
\begin{itemize}
\item It improves the realism of the simulations of cellular systems (except perhaps for systems in pathological situations): no fragmentation or nucleation occurs, and cells stay simply connected.
\item For a same simulation temperature, it is faster than the standard algorithm used in CPM simulations: the time spent to test the local connectivity of the cells is largely offset by the time spared by not doing moves that induce fragmentation.
\item It restores detailed balance. As a consequence, the long-term stage and the statistics of configurations do not depend on the specific chosen acceptance rate, nor on the chosen path in the simulation temperature space: it depends only on the final temperature once the thermal equilibrium is reached. 
%
%
%
\item It works for all simulation temperatures. Hence, when interested in the long-term stage of the simulations, we can (temporarily) increase the simulation temperature to converge more rapidly.
\item Its implementation is much easier than those of parallel algorithms.
\end{itemize}


Note that if this algorithm is intended to be used for prohibiting cell fragmentation, and not so for preserving the simple connectivity of the cells or the detailed balance \revision{criterion}, the local connectivity test on the target cell (step 4c.) can be skipped.

\begin{figure}[htb]
\centering
\subfigure[]{
    \includegraphics[width=0.2\textwidth]{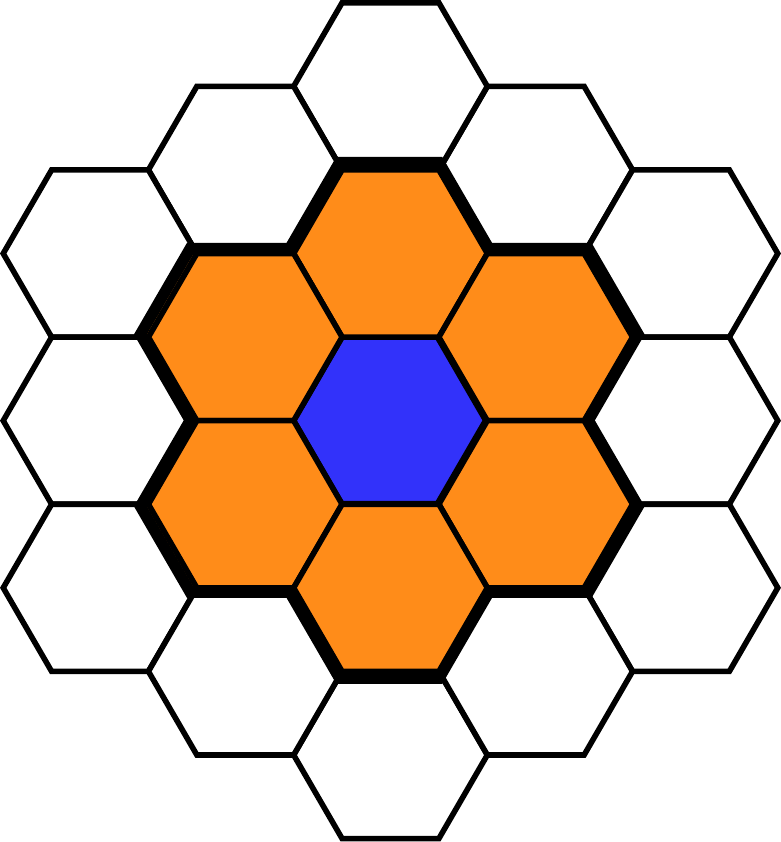}
    \label{hexagonal_lattice}}
\hfill
\subfigure[]{
    \includegraphics[width=0.22\textwidth]{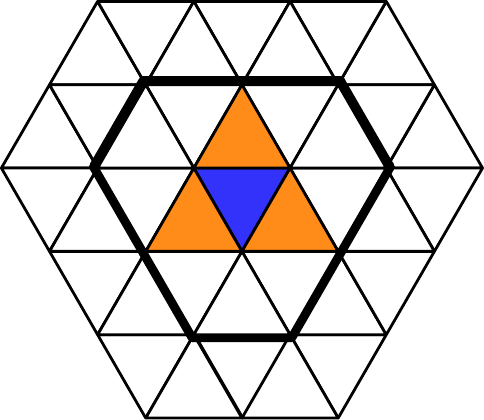}
    \label{triangular_lattice}}
\hfill
\subfigure[]{
    \includegraphics[width=0.22\textwidth]{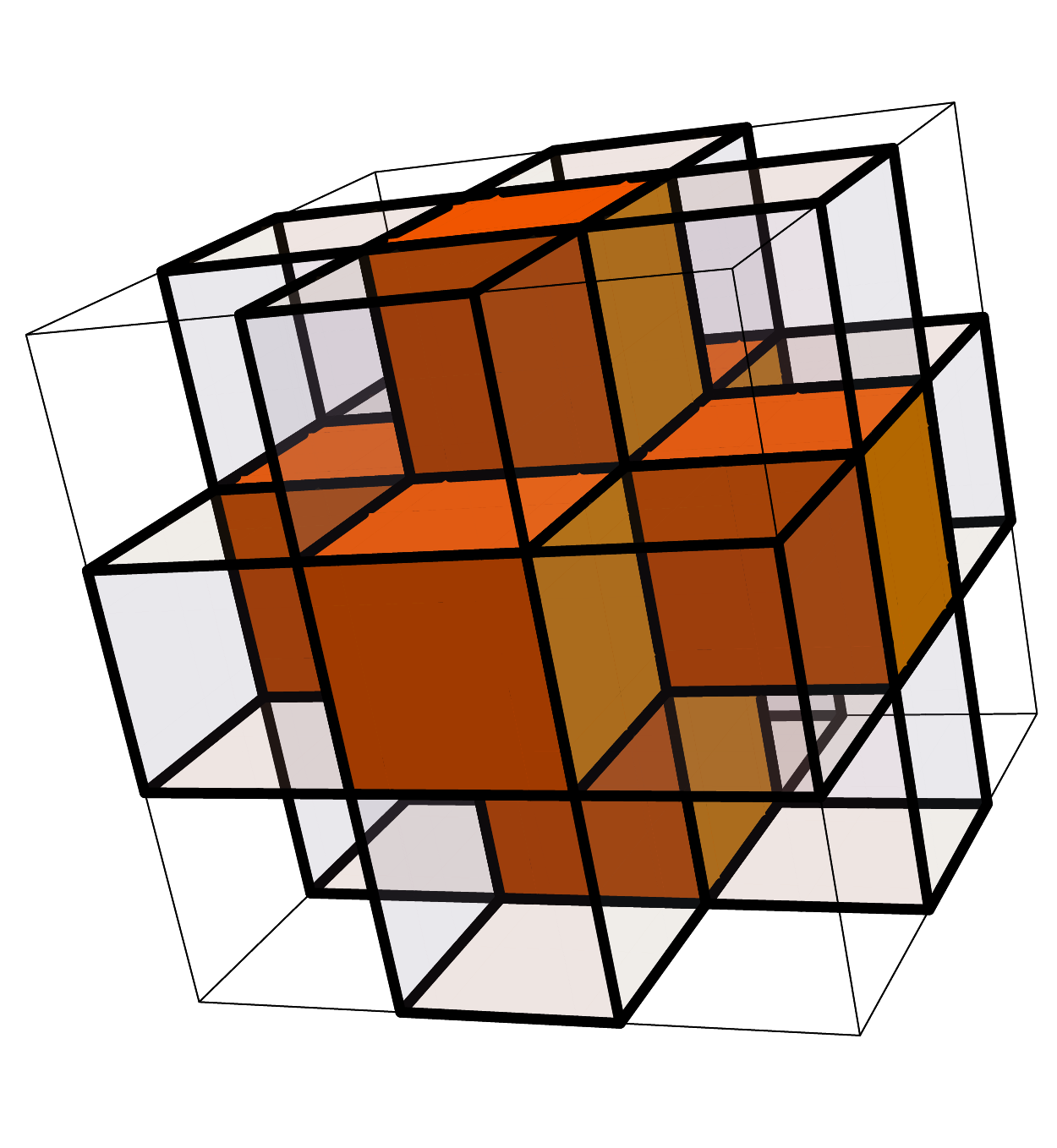}
    \label{cubic_lattice}}
\caption{\revision{Best choices for the adjacency neighborhood, target neighborhood, and local connectivity domain in (a) hexagonal lattice; (b) triangular lattice; (c) cubic lattice. Adjacency and target neighbourhoods are equal and shown in orange. The local connectivity domain is delimited by bold edges (and shadowed surface for the 3D lattice).
}
}
\label{fig:lattices}
\end{figure}

Although we focused primarily on the 2D square lattice, the algorithm can be adapted without difficulty to other 2D or 3D lattices. At 2D, alternatives are the hexagonal and triangular lattices \revision{(see Figs \ref{hexagonal_lattice} and \ref{triangular_lattice})}. Best choices for the values of $\Na$ ($=\Nt$) and $\D_c$ are generalized as follows: adjacency neighborhood is defined as the set of lattice sites that are side-adjacent to the central site (hence composed of 6 sites for the hexagonal lattice, and 3 sites for the triangular lattice). Indeed, this is the minimal number of lattice sites required to respect the lattice symmetry. Local connectivity domain $\D_c$ is defined as the set of lattice sites that are side- or corner-adjacent to the central site (hence composed of the same 6 sites than $\Na$ for the hexagonal lattice, and 12 sites for the triangular lattice).
For the extension to the 3D cubic lattice, $\Na$ and $\Nt$ are composed of the 6 face-adjacent sites, and $\D_c$ of the 12 face- or edge-adjacent sites \revision{(see Fig. \ref{cubic_lattice})}.

The modifications brought by our algorithm also allow to use other families of Monte-Carlo algorithms. In particular, one can think of \textit{single-spin-exchange} algorithms, like the Kawasaki algorithm, as opposed to the Metropolis or Heat bath algorithms which are \textit{single-spin-flip} algorithms. In the Kawasaki algorithm, two lattice sites exchange their values, conserving the number of sites (\ie area) of every cell, what makes the compressive energy term in the Potts Hamiltonian (\ref{Hamiltonian}) unnecessary. Kawasaki algorithm is commonly used in interface problems \cite{newman_monte_1999}. As with Metropolis, the algorithm must be adapted to prevent spontaneous nucleation: only exchange of neighboring sites' values (local value-exchange) must be allowed. With the MMA, a series of local value-exchanges would still result in a lone heterogeneous site in a region of differing site values. Because cell areas are conserved, the lone heterogeneous site would persist for much longer time than it would with Metropolis. Presumably this is the reason why the Kawasaki algorithm has never been used in CPM simulations, to our knowledge. Kawasaki algorithm can be safely used with the CA, thanks to the cell connectivity test.

It can also be noticed that the significant difference in kinetics of cell sorting produced by the two algorithms may reduce the discrepancy observed so far between experimental and CPM  sorting processes \cite{Nakajima_2011, Beysens_2000}. Further investigation are currently conducted in order to confirm (or infirm) this hypothesis.
 
\begin{figure}[htb]
\centering
\includegraphics[width = 0.4\textwidth]{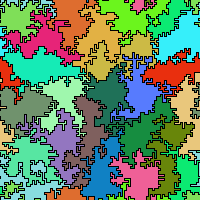}
\caption{Close-up of the cellular pattern obtained with the CA at $T=130$: cells have contorted structures. Each cell is coded with a different color to facilitate their visualization.
}
\label{fig:dendritic}
\end{figure}

\revision{Finally, it is worth mentioning that our algorithm may open up new fields of application for CPM simulations: at higher simulation temperature, prohibition of fragmentation leads to the formation of dendritic cells, as shown in Fig. (\ref{fig:dendritic}). Combined with anisotropic neighborhoods, prohibition of fragmentation may also facilitate simulating of elongated cells.}

\section*{Acknowledgements}
We would like to thank F. Graner for his support and insightful discussions.

\appendix
\section*{Appendix}
\begin{figure}[htb]
\begin{center}
\includegraphics[width=0.65\columnwidth]{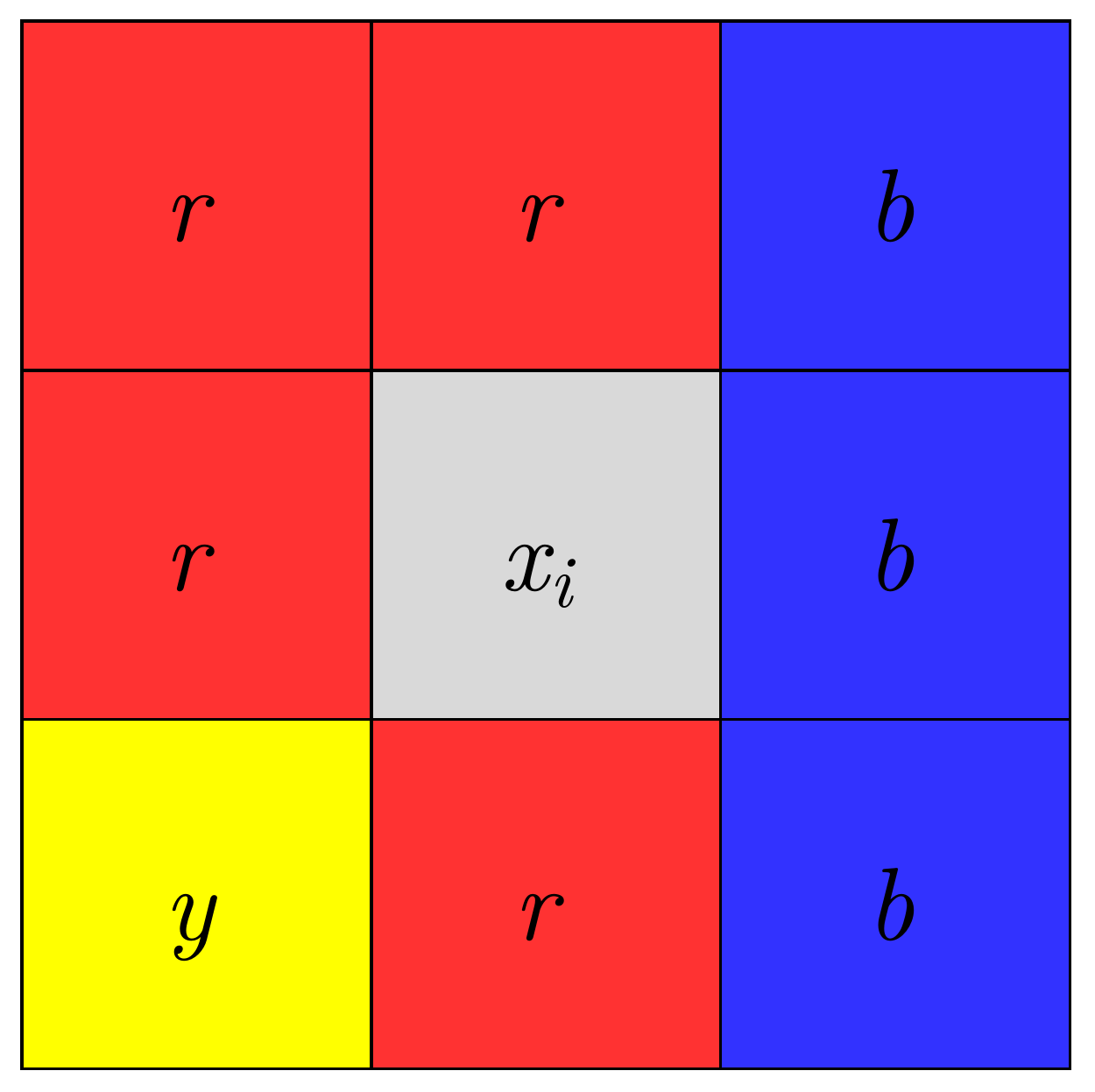}
\caption{Examples of target neighborhoods for which detailed balance condition is not satisfied with the target value selection used within the MMA (step (2b.)): either with a target neighborhood equal to the Von Neumann or Moore neighborhood, the site value $r$ has more chance to be selected than site values $b$ or $y$.
}
\label{fig_detailed_balance}
\end{center}
\end{figure}

The Modified Metropolis Algorithm (MMA) commonly used in CPM simulation differs from the standard Metropolis algorithm in the selection of target values (step (2b.) \textit{vs} step (2.)): in standard Metropolis, the target value is randomly chosen from any of the $Q$ possible values, without bias, where $Q$ denotes the number of cells. In the MMA, the target value is chosen from the list of neighbouring sites, without bias. As a consequence, the selection probability does not satisfy $g(x_i:\sigma\rightarrow\sigma^\prime)= g(x_i:\sigma^\prime\rightarrow\sigma)$ anymore, and the condition of detailed balance [Eq. (\ref{detailed_balance})] is not satisfied. This is exemplified with the neighborhood shown in Fig. \ref{fig_detailed_balance}: if we choose for the target neighborhood $\Nt$ (see Sect. \ref{Modified_algo}) the 4 side-adjacent lattice sites (Von Neumann neighborhood), we have: $g(x_i:r\rightarrow b)= 1/(4N)$, while $g(x_i:b\rightarrow r)= 3/(4N)$. If $\Nt$ is composed of the 8 side- or corner-adjacent sites (Moore neighborhood): $g(x_i:r\rightarrow b)= 3/(8N)$, while $g(x_i:b\rightarrow r)= 4/(8N)$.
One way to restore detailed balance would be to weight acceptance ratios to recover condition of detailed balance: the original acceptance ratio $\mathcal{A}(x_i:\sigma\rightarrow\sigma^\prime)$ is replaced with $\mathcal{A}(x_i:\sigma\rightarrow\sigma^\prime)/z(\sigma^\prime)$, where $z(\sigma^\prime)$ is the number of sites, within the target neighborhood, which have the specific value $\sigma^\prime$ \cite{Zajac_2002}. However, this is not the best way to restore detailed balance, as we should keep acceptance ratios as large as possible to have an efficient algorithm. Instead, we choose to modify the selection probability by replacing step (2b.) (Sect. \ref{Modified_algo}) with:
\begin{enumerate}[label=\arabic*c.]
\setcounter{enumi}{1}
\item Randomly select a value from those present in the target neighborhood $\Nt$. Call this the target value. Let $\sigma^\prime$ be its value.
\end{enumerate}
This restores the equality $g(x_i:\sigma\rightarrow\sigma^\prime)= g(x_i:\sigma^\prime\rightarrow\sigma)$. In the example of Fig. \ref{fig_detailed_balance}, the number of possible target values is $2$ for a Von Neumann target neighborhood, and $3$ for a Moore neighborhood. Thus, $g(x_i:r\rightarrow b)= g(x_i:b\rightarrow r)=1/(2N)$ in the first case, and $=1/(3N)$ in the second case.

In comparison with step (2b.), step (2c.) is somehow a more straightforward adaptation of step (2.): in step (2c.) the target value is drawn from among the set of values within $\Nt$, while in step (2.) the target value is drawn from among the $Q$ values within the whole pattern.
However, compared to step (2b.), step (2c.) would enhance persistence of fragments,
if used without the modification brought by our algorithm to prevent bubble fragmentation. We presume that is the reason why target value selection (2b.) prevails in most CPM simulations.  
\linebreak
\bibliographystyle{elsarticle-num}
\bibliography{biblio}

\begin{thebibliography}{10}
\expandafter\ifx\csname url\endcsname\relax
  \def\url#1{\texttt{#1}}\fi
\expandafter\ifx\csname urlprefix\endcsname\relax\def\urlprefix{URL }\fi
\expandafter\ifx\csname href\endcsname\relax
  \def\href#1#2{#2} \def\path#1{#1}\fi

\bibitem{graner_simulation_1992}
F.~Graner, J.~A. Glazier, Simulation of biological cell sorting using a
  two-dimensional extended {Potts} model, Physical Review Letters 69~(13)
  (1992) 2013--2016.

\bibitem{glazier_simulation_1993}
J.~Glazier, F.~Graner, Simulation of the differential adhesion driven
  rearrangement of biological cells, Physical Review E 47~(3) (1993)
  2128--2154.

\bibitem{glazier_magnetization_2007}
J.~A. Glazier, A.~Balter, N.~J. Popławski, Magnetization to {Morphogenesis}:
  {A} {Brief} {History} of the {Glazier}-{Graner}-{Hogeweg} {Model}, in: D.~A.
  R.~A. Anderson, P.~M. A.~J. Chaplain, D.~K.~A. Rejniak (Eds.),
  Single-{Cell}-{Based} {Models} in {Biology} and {Medicine}, Mathematics and
  {Biosciences} in {Interaction}, Birkh\"auser Basel, 2007, pp. 79--106.

\bibitem{Steinberg_1963}
M.~S. Steinberg, Reconstruction of tissues by dissociated cells, Science
  141~(3579) (1963) 401--408.

\bibitem{mombach_quantitative_1995}
J.~C. Mombach, J.~A. Glazier, R.~C. Raphael, M.~Zajac, Quantitative comparison
  between differential adhesion models and cell sorting in the presence and
  absence of fluctuations, Physical Review Letters 75~(11) (1995) 2244.

\bibitem{ouchi_improving_2003}
N.~B. Ouchi, J.~A. Glazier, J.-P. Rieu, A.~Upadhyaya, Y.~Sawada, Improving the
  realism of the cellular {Potts} model in simulations of biological cells,
  Physica A: Statistical Mechanics and its Applications 329~(3-4) (2003)
  451--458.

\bibitem{magno_biophysical_2015}
R.~Magno, V.~A. Grieneisen, A.~F. Marée, The biophysical nature of cells:
  potential cell behaviours revealed by analytical and computational studies of
  cell surface mechanics, BMC Biophysics 8~(1).

\bibitem{mombach_rounding_2005}
J.~C.~M. Mombach, D.~Robert, F.~Graner, G.~Gillet, G.~L. Thomas, M.~Idiart,
  J.-P. Rieu, Rounding of aggregates of biological cells: {Experiments} and
  simulations, Physica A: Statistical Mechanics and its Applications 352~(2-4)
  (2005) 525--534, arXiv: cond-mat/0411647.

\bibitem{kafer_cell_2007}
J.~Käfer, T.~Hayashi, A.~F.~M. Marée, R.~W. Carthew, F.~Graner, Cell adhesion
  and cortex contractility determine cell patterning in the {Drosophilaretina},
  Proceedings of the National Academy of Sciences 104~(47) (2007) 18549--18554.

\bibitem{oates_quantitative_2009}
A.~C. Oates, N.~Gorfinkiel, M.~González-Gaitán, C.-P. Heisenberg,
  Quantitative approaches in developmental biology, Nature Reviews Genetics
  10~(8) (2009) 517--530.

\bibitem{ariotti_tissue-resident_2012}
S.~Ariotti, J.~B. Beltman, G.~Chodaczek, M.~E. Hoekstra, A.~E.~v. Beek,
  R.~Gomez-Eerland, L.~Ritsma, J.~v. Rheenen, A.~F.~M. Marée, T.~Zal, R.~J.~d.
  Boer, J.~B. A.~G. Haanen, T.~N. Schumacher, Tissue-resident memory {CD}8+ {T}
  cells continuously patrol skin epithelia to quickly recognize local antigen,
  Proceedings of the National Academy of Sciences 109~(48) (2012) 19739--19744.

\bibitem{fortuna_growth_2012}
I.~Fortuna, G.~L. Thomas, R.~M.~C. de~Almeida, F.~Graner, Growth {Laws} and
  {Self}-{Similar} {Growth} {Regimes} of {Coarsening} {Two}-{Dimensional}
  {Foams}: {Transition} from {Dry} to {Wet} {Limits}, Physical Review Letters
  108~(24) (2012) 248301.

\bibitem{Albert2014}
P.~J. {Albert}, U.~S. {Schwarz}, Dynamics of cell shape and forces on
  micropatterned substrates predicted by a cellular potts model, Biophysical
  Journal 106~(11) (2014) 2340--2352.

\bibitem{Kabla3268}
A.~J. Kabla, Collective cell migration: leadership, invasion and segregation,
  Journal of The Royal Society Interface 9~(77) (2012) 3268--3278.

\bibitem{Hallou2016}
A.~{Hallou}, J.~{Jennings}, A.~{Kabla}, {Cancer Metastasis: Collective Invasion
  in Heterogeneous Multicellular Systems}, ArXiv e-prints\href
  {http://arxiv.org/abs/1501.00065} {\path{arXiv:1501.00065}}.

\bibitem{gusatto_efficient_2005}
E.~Gusatto, J.~C.~M. Mombach, F.~P. Cercato, G.~H. Cavalheiro, An efficient
  parallel algorithm to evolve simulations of the cellular potts model,
  Parallel Processing Letters 15~(01n02) (2005) 199--208.

\bibitem{chen_parallel_2007}
N.~Chen, J.~A. Glazier, J.~A. Izaguirre, M.~S. Alber, A parallel implementation
  of the {Cellular} {Potts} {Model} for simulation of cell-based morphogenesis,
  Computer Physics Communications 176~(11-12) (2007) 670--681.

\bibitem{tapia_parallelizing_2011}
J.~J. Tapia, R.~M. D'Souza, Parallelizing the {Cellular} {Potts} {Model} on
  graphics processing units, Computer Physics Communications 182~(4) (2011)
  857--865.
\newblock \href {http://dx.doi.org/10.1016/j.cpc.2010.12.011}
  {\path{doi:10.1016/j.cpc.2010.12.011}}.

\bibitem{Zajac_2002}
M.~{Zajac}, {Modeling convergent extension by way of anisotropic differential
  adhesion}, Ph.D. thesis, University of Notre Dame (2002).

\bibitem{voss-bohme_multi-scale_2012}
A.~Voss-B\"ohme, Multi-scale modeling in morphogenesis: a critical analysis of
  the cellular {Potts} model, PloS one 7~(9) (2012) e42852.

\bibitem{newman_monte_1999}
M.~Newman, G.~Barkema, Monte Carlo Methods in Statistical Physics, Clarendon
  Press ; Oxford University Press, Oxford : New York, 1999.

\bibitem{Durand_PRL}
M.~Durand, H.~A. Stone, Relaxation time of the topological $t1$ process in a
  two-dimensional foam, Phys. Rev. Lett. 97 (2006) 226101.

\bibitem{Durand_EPJE_2015}
M.~{Durand}, Statistical mechanics of two-dimensional foams: Physical
  foundations of the model, Eur. Phys. J. E 38~(12) (2015) 137.

\bibitem{Nakajima_2011}
A.~Nakajima, S.~Ishihara, Kinetics of the cellular potts model revisited, New
  Journal of Physics 13~(3) (2011) 033035.

\bibitem{Beysens_2000}
D.~A. Beysens, G.~Forgacs, J.~A. Glazier, Cell sorting is analogous to phase
  ordering in fluids, Proceedings of the National Academy of Sciences 97~(17)
  (2000) 9467--9471.

\end{thebibliography}

\end{document}